\documentclass[preprint,showpacs,preprintnumbers,amsmath,amssymb]{revtex4-2}
\usepackage{bm}% bold math
\usepackage{amsmath}
\usepackage{amssymb}
\usepackage{amsfonts}
\usepackage{graphicx}
\begin{document}

\title{ \textbf{
    Investigation of Entanglement in Pure Final Polarization States
    from Neutron–Deuteron Elastic Scattering and Exclusive Deuteron Breakup
} }

\author{H.~Wita{\l}a}
\affiliation{M. Smoluchowski Institute of Physics, 
Faculty of Physics, Astronomy and Applied Computer Science,
Jagiellonian University, PL-30348 Krak\'ow, Poland}

%\author{J.~Golak}
%\affiliation{M. Smoluchowski Institute of Physics, 
%Faculty of Physics, Astronomy and Applied Computer Science,
%Jagiellonian University, PL-30348 Krak\'ow, Poland}

%\author{R.~Skibi\'nski}
%\affiliation{M. Smoluchowski Institute of Physics, 
%Faculty of Physics, Astronomy and Applied Computer Science,
%Jagiellonian University, PL-30348 Krak\'ow, Poland}

\date{\today}

\begin{abstract}
  We investigate the pure polarization states of the outgoing neutron–deuteron
  ($nd$) pair in elastic polarized neutron–polarized deuteron scattering,
  as well as the pure polarization states of the three free nucleons ($nnp$)
  produced in the corresponding deuteron breakup reaction. Our aim is
  to provide clear evidence of entanglement in their spin degrees of freedom.
  These final states can be generated from pure spin states of the incoming
  $nd$ system, where both the neutron and the deuteron are strongly
  polarized by maximizing the occupation of a specific magnetic substate.

  To fully characterize the final configurations, we compute
  the corresponding spin density matrices using the high-precision
  CD-Bonn nucleon–nucleon potential. Among these pure final spin states,
  we identify strongly entangled, Bell-like states that include at most
  small admixtures of components diminishing entanglement.
\end{abstract}

%\pacs{21.45.-v, 21.45.Bc, 25.10.+s, 25.40.Cm}

\maketitle \setcounter{page}{1}

\section{Introduction}
\label{intro}

Studies of reactions with polarized incoming particles provide a unique
way to systematically generate the final spin states of outgoing nucleons
by varying the polarizations of the reactants. In the simplest and most
extensively studied nuclear processes over the past decades—both
experimentally and theoretically—namely neutron–proton ($np$) and
proton–proton ($pp$) scattering—the outgoing nucleons are polarized
even when the incoming state is unpolarized. This induced polarization
originates from the spin–orbit term in the nucleon–nucleon ($NN$) potential.

Measuring the cross section with a single polarized nucleon yields the
analyzing powers, which correspond to the induced polarizations in
unpolarized $NN$ scattering. The outgoing polarizations also acquire
additional contributions, providing the corresponding single-spin transfer
coefficients. Polarizing both beam and target allows determination of
the polarization transfer from a doubly spin-polarized initial
state~\cite{ohlsen1972,witnp2024}, which differs from single-spin transfers.
Each outgoing nucleon can thus acquire four contributions: the fixed
induced polarization, two single-spin transfers, and the transfer from
the doubly polarized initial state, all adjustable via the initial
polarizations.

In standard nuclear scattering experiments, the beam and target
polarizations are prepared independently, with the initial $np$ density
matrix given by the cross-product of the beam and target matrices. Such
a state is unentangled and, for partially polarized particles, also mixed,
resulting in final spin states that are predominantly statistical mixtures.

The study of nuclear entanglement and its experimental signatures in
final nucleon pairs has recently attracted growing interest, motivating
investigations in the $np$ system. Entanglement generation in $np$
scattering has been extensively explored;
see, e.g.,~\cite{beane2019,beane2021,bai2022,bai2023,liu2023,kirch2,bai2024}.
Calculations of the entanglement power—which quantifies the entanglement
induced by the $S$-matrix acting on an initially unentangled state—have
provided insight into the role of the $NN$ interaction in generating spin
correlations.

In Ref.~\cite{witnp2024}, this problem was analyzed using the transition
operator $T$, which directly yields the outgoing neutron and proton
polarizations and their spin correlations, fully determining the spin
density matrix of the final $np$ pair. The analysis revealed increasing
entanglement with rising $np$ energy. For maximally polarized initial states,
which are pure, several Bell-like states were identified among the pure
final outcomes.

The present study extends this approach to the $nd$ system, investigating
entanglement in the final pure polarization states of $nd$ elastic scattering
and the exclusive deuteron breakup reaction. By selecting incoming neutron
and deuteron polarizations that produce an unentangled but pure initial $nd$
state, the final states remain pure. Among these, entangled Bell-like
states appear in both elastic scattering and breakup, providing a
controlled setting to explore entanglement generation in three-nucleon systems.

In Sec.~\ref{form}, we briefly outline the three-nucleon Faddeev formalism,
which provides the transition amplitudes for $nd$ elastic scattering and
deuteron breakup, along with the essentials of the spin formalism. Results
on entanglement, including examples of Bell-like states identified in
elastic $nd$ scattering and in the exclusive deuteron breakup, are presented
in Secs.~\ref{elas} and~\ref{breakup}, respectively.
Finally, Sec.~\ref{sumary} contains a summary and conclusions.

\section{Formalism}
\label{form}

Determination of the spin density matrices for the final states requires
the transition amplitudes for elastic $nd$ scattering and breakup.
For the reader's convenience, we briefly outline the main points of
the three-nucleon (3N) Faddeev formalism, which describes both processes.
For details of the formalism and its numerical implementation,
we refer to~\cite{glo96,wit88,hub97,book}.

Nucleon–deuteron ($Nd$) scattering, with nucleons interacting solely via
the nucleon–nucleon ($NN$) interaction $v_{NN}$, is described in terms of
the breakup operator $T$, which satisfies the Faddeev-type
integral equation~\cite{glo96,wit88,hub97}
\begin{eqnarray}
T\vert \phi \rangle  &=& t P \vert \phi \rangle +
t P G_0 T \vert \phi \rangle  \, .
\label{eq1a}
\end{eqnarray}
The two-nucleon (2N) $t$-matrix, $t$, is the solution of the
Lippmann-Schwinger equation with the interaction
$v_{NN}$. The permutation operator $P=P_{12}P_{23} +
P_{13}P_{23}$ is expressed in terms of the transposition operators 
$P_{ij}$, which interchange  nucleons $i$ and $j$.  The initial state 
$\vert \phi \rangle = \vert \vec {q}_0 \rangle \vert \phi_d \rangle$
describes the free motion of the nucleon  and the deuteron 
  with  relative momentum
  $\vec {q}_0$  and contains the internal deuteron wave function
  $\vert \phi_d \rangle$. Finally,  
  $G_0$ denotes the free three-body resolvent.

 The amplitude for elastic scattering leading to the 
 final Nd state $\vert \phi ' \rangle$ is then given by~\cite{glo96,hub97}
\begin{eqnarray}
\langle \phi' \vert U \vert \phi \rangle &=& \langle \phi' 
\vert PG_0^{-1} \vert 
\phi \rangle  
 + 
\langle \phi' \vert PT \vert \phi \rangle ~,
\label{eq3}
\end{eqnarray}
while the  amplitude for the breakup reaction is expressed as
\begin{eqnarray}
\langle  \vec p \ \vec q \ \vert U_0 \vert \phi \rangle &=&\langle 
 \vec p \  \vec q \  \vert  (1 + P)T\vert
 \phi \rangle ,
\label{eq3_br}
\end{eqnarray}
where the free  breakup channel state $\vert  \ \vec p \ \vec q \ \rangle $
is defined in terms of the  Jacobi (relative) momenta $\vec p$
and $\vec q$. 

We solve Eq.~(\ref{eq1a}) in the momentum-space partial-wave basis
$\vert p q \alpha \rangle$, which is specified by the magnitudes of
the Jacobi momenta $p$ and $q$ and a set of discrete quantum numbers
$\alpha$. These include the 2N subsystem spin, isospin, orbital, and
total angular momenta $s$, $t$, $l$, and $j$, as well as the orbital
and total angular momenta of the spectator nucleon relative to the center
of mass (c.m.) of the 2N subsystem, $\lambda$ and $I$:
\begin{eqnarray}
\vert p q \alpha \rangle \equiv \vert p q (ls)j (\lambda \frac {1} {2})I (jI)J
  (t \frac {1} {2})T \rangle ~.
\label{eq4a}
\end{eqnarray}
The total 2N subsystem and spectator nucleon angular momenta, $j$ and $I$,
as well as the isospins $t$ and $\frac{1}{2}$, are finally coupled to
the total angular momentum $J$ and isospin $T$ of the 3N system, respectively.
In practice, a converged solution of Eq.~(\ref{eq1a}) using a partial-wave
decomposition in momentum space at a given energy $E$ requires including
all 3N partial-wave states up to 2N angular momentum $j_{\rm max} = 5$
and 3N angular momentum $J_{\rm max} = \frac{25}{2}$.

  In the following, we briefly summarize the main points of the spin formalism
  essential for elastic $nd$ scattering, $\vec d(\vec n,\vec n)\vec d$,
  and the deuteron breakup reaction,
  $\vec d(\vec n,\vec N_1 \vec N_2)\vec N_3$, with polarized incoming
  neutron and deuteron and outgoing products in appropriate final spin states.
  When both the incoming neutron and deuteron are polarized,
  the polarizations of the outgoing particles contain all contributions:
  not only the induced polarization and single-spin transfers from
  the neutron or deuteron, but also a component arising from
  the doubly polarized initial state~\cite{doubl_poltrans}.

  The applied notation for the deuteron breakup reaction emphasizes its
  exclusive character, with the momenta of nucleons $N_1$ and $N_2$ measured
  in coincidence: the detector for nucleon $N_1$ is placed at laboratory
  angles $(\theta_1,\phi_1)$ and for nucleon $N_2$ at $(\theta_2,\phi_2)$,
  thereby providing a complete kinematical specification of the
  configuration. Both processes are described by a transition
  amplitude $T^{{m_{in}}}_{{m_{out}}}$, equal to $U$ of Eq.~(\ref{eq3})
  for elastic scattering, or $U_0$ of Eq.~(\ref{eq3_br}) for breakup,
  with the corresponding sets of spin projections ${m_{in}}$ and ${m_{out}}$
  for the incoming and outgoing particles, respectively. For further
  details of the spin formalism, we refer to~\cite{ohlsen1972,simonius1973}.

  Polarizations are always described using right-handed Cartesian coordinate
  systems defined according to the Madison Convention~\cite{Madison1971}.

  In a standard scattering experiment, the density matrix of the initial state,
  $\rho^{\rm in}$, is expressed in terms of the polarizations of the
  incoming particles. For the $nd$ initial state, these are the neutron
  vector polarization $p_i^n$, and the deuteron vector and tensor
  polarizations, $p_i^d$ and $p_{ij}$,
  respectively~\cite{ohlsen1972,simonius1973}:
\begin{eqnarray}
  \rho^{in} &=& \rho^{in}_n  \otimes \rho^{in}_d = 
      [ \frac {1} {2} ( I + \sum_{i=1}^3 p_i^n \sigma_i  ] \otimes  \
 [ \frac {1} {3} ( I +  \frac {3} {2}  \sum_{i=1}^3 p_i^d P_i \cr
      &+&  \frac {2} {3} ( p_{xy} P_{xy} + p_{yz} P_{yz} + p_{xz} P_{xz} )
      +  \frac {1} {6} (p_{xx}-p_{yy})(P_{xx}-P_{yy})
      +  \frac {1} {2} p_{zz} P_{zz} ] ~.
\label{eq1_a0}
\end{eqnarray}
Here, $\sigma_i$ are the standard Pauli matrices, and $P_i$ and $P_{ij}$
denote the set of spin-1 operators defined in~\cite{ohlsen1972}.
The polarizations are given by
$$p_i^n \equiv <\sigma_i>_{\rho^{in}_n} = {\text{Tr}}( \rho^{in}_n \sigma_i ),~~
p_i^d \equiv <P_i>_{\rho^{in}_d} = {\text{Tr}}( \rho^{in}_d P_i ),~~
p_{ij} \equiv <P_{ij}>_{\rho^{in}_d} = {\text{Tr}}( \rho^{in}_d P_{ij} ),$$
where $\langle O \rangle_{\rho}$ denotes the expectation value of the operator
$O$ in the state described by the density matrix $\rho$.

The density matrix $\rho^{\rm in}$, together with the transition operator
$T$, determines the density matrix of the final state, $\rho^{\rm out}$,
normalized to unit trace:
\begin{eqnarray}
  \rho^{out} &=& \frac {T \rho^{in} T^{\dagger}}  {\text{Tr}
      (T \rho^{in} T^{\dagger}) } ~.
\label{eq2_a0}
\end{eqnarray}

Spin states described by the density matrices $\rho^{\rm in}_n$,
$\rho^{\rm in}_d$, and $\rho^{\rm in}$ of Eq.~(\ref{eq1_a0}) are, in
general, statistical mixtures of states. Only for specific values of
the polarizations do these matrices correspond to a normalized pure
spin state, $|\psi^{\rm spin} \rangle$. The corresponding density
matrix $\rho$ is then given by:
\begin{eqnarray}
 \rho &=& |\psi^{spin} \rangle  \langle \psi^{spin} | ~,
\label{eq14_a0}
\end{eqnarray}
it has unit trace, $\mathrm{Tr}(\rho) = 1$, and also satisfies
the pure-state idempotency condition:
\begin{eqnarray}
  (\rho)^2 &=& \rho ~.
\label{eq15_a0}
\end{eqnarray}

It is evident that the properties of the final spin states in $nd$ scattering
and breakup are largely determined by the polarization state of the
incoming particles, which is taken in the form of Eq.~(\ref{eq1_a0}), with
the neutron beam and deuteron target polarizations prepared entirely
independently. Such a state is clearly not entangled and, for partially
polarized particles, also impure, leading to final spin states that are
predominantly statistical mixtures.

Since we are interested in signals of entanglement in the final pure
polarization states of $nd$ elastic scattering and the deuteron breakup
reaction, we aim to prepare the incoming $nd$ system in a pure spin state.
When the initial spin state is pure, the final state must also be pure.
Specifically, a pure initial state $|\psi\rangle$, normalized to unity,
leads to an idempotent density matrix,
$\rho^{\rm in} = |\psi\rangle \langle \psi|$, with unit trace.
The corresponding final state is then described by the density
matrix $\rho^{\rm out}$:
\begin{eqnarray}
  \rho^{out} &=& \frac {T |\psi\rangle  \langle \psi | T^{\dagger}}
      {\text{Tr} (T |\psi\rangle 
    \langle \psi | T^{\dagger}  )}
\label{new3} ~,
\end{eqnarray}
which satisfies the idempotency condition~(\ref{eq15_a0})~\cite{witnp2024},
and the corresponding pure final state is given by:
\begin{eqnarray}
\   |\psi^{out} \rangle &=& \frac {T |\psi\rangle } {[\text{Tr} (T |\psi\rangle
\langle \psi | T^{\dagger} )]^{\frac {1} {2} }} ~.
\label{new6} 
\end{eqnarray}

In the following, we restrict ourselves to axially symmetric pure incoming
polarized states of neutrons and deuterons, with the symmetry axis chosen
along the $y$ direction. The neutron and deuteron spin density matrices
then take the forms
$$ \rho^{in}_n =
  \frac {1} {2} ( I +     p_y^n \sigma_y ),~~~  \rho^{in}_d =
  \frac {1} {3} ( I +  \frac {3} {2}   p_y^d P_y +
  \frac {1} {2} p_{yy} P_{yy} ), $$
with polarization values $p_y^n$, $p_y^d$, and $p_{yy}$ determined by the
maximal occupation numbers of the various magnetic
substates~\cite{ohlsen1972}. When the neutron states
${m'}_n = \pm \frac{1}{2}$ and deuteron states ${m'}_d = \pm 1$ are
maximally populated, one obtains $p_y^n = \pm 1$, $p_y^d = \pm 1$,
and $p_{yy} = +1$, whereas maximal population of the ${m'}_d = 0$ state
yields $p_y^d = 0$ and $p_{yy} = -2$. The prime on the spin projection
$m'$ emphasizes that the quantization axis is chosen along the $y$ direction.

By combining the two values of the neutron polarization, $p_y^n = \pm 1$,
which correspond to pure incoming neutron spin states, with the three
values of the deuteron polarization, $p_y^d = \pm 1$ and $p_y^d = 0$, which
correspond to pure incoming deuteron spin states, one obtains six
combinations of $(p_y^n, p_y^d, p_{yy})$, listed in the first column of
Table~\ref{tab1}. Each combination defines a pure and unentangled
neutron–deuteron cross-product state, from which pure final states
are generated in both elastic scattering and breakup. It is this set of
resulting pure final states that we examine for signatures of entanglement.

Let us now focus on these specific pure initial states for different
combinations of neutron and deuteron polarizations. The state of the
neutron with spin projection $m$ along the $y$-axis can be obtained
from the states with spin projections ${\bar m} = \pm \frac{1}{2}$
along the $z$-axis by a rotation
\[
| \frac {1} {2} m \rangle ^y =
\sum\limits_{\bar m}
D^{\frac {1} {2} }_{{\bar m}m} (\alpha \beta \gamma)
| \frac {1} {2} \bar m \rangle ^z 
\]
with the Euler angles $\alpha = \frac{\pi}{2}$, $\beta = \frac{\pi}{2}$,
and $\gamma = 0$~\cite{brinksatch}, and similarly for the deuteron. 
This then leads, for the neutron, to
\begin{eqnarray}
|\frac {1} {2},+\frac {1} {2} \rangle ^y &=&  
\frac {\sqrt{2}} {2}e^{-i\frac {\pi} {4}} \left(+|\frac {1} {2},+\frac {1} {2} \rangle  +
i\, |\frac {1} {2},-\frac {1} {2} \rangle  \right) \label{new1} \, , \\
|\frac {1} {2},-\frac {1} {2} \rangle ^y &=& 
\frac {\sqrt{2}} {2}e^{-i\frac {\pi} {4}} \left(-|\frac {1} {2},+\frac {1} {2} \rangle  +
i\, |\frac {1} {2},-\frac {1} {2} \rangle  \right) \label{new2} ~,
\end{eqnarray}
and, for the deuteron, to
\begin{eqnarray}
|1,+1 \rangle ^y &=&  
  -\frac {i} {2} \, |1,+1 \rangle + \frac {1} {\sqrt{2}}\, |1,~~0 \rangle 
 + \frac {i} {2} \, |1,-1 \rangle   \label{neww1} \, , \\
|1,~~0 \rangle ^y &=& 
 \frac {i} {\sqrt{2}}\, |1,+1 \rangle  ~~~~~~~~~~~~~~~  +
 \frac {i} {\sqrt{2}}\, |1,-1 \rangle   \label{neww2} \, , \\
|1,-1 \rangle ^y &=& 
 -\frac {i} {2} \, |1,+1 \rangle -\frac {1} {\sqrt{2}}\, |1,~~0 \rangle  + 
\frac {i} {2} \, |1,-1 \rangle   \label{neww3} ~.
\end{eqnarray}

Thus, neglecting the irrelevant overall factor $e^{-i \frac{\pi}{2}}$ in
Eqs.~(\ref{new1}) and (\ref{new2}), the six pure initial states for
different combinations of the neutron polarization, $p_y^n = \pm 1$,
and the deuteron polarizations, $p_y^d = \pm 1$, $p_{yy} = +1$,
and $p_y^d = 0$, $p_{yy} = -2$, are given by:
\begin{eqnarray}
  |\psi_{nd}^{in,spin} \rangle &=& \alpha_{+\frac {1} {2} +1}
  | {+\frac {1} {2}} {+1}\rangle
  + \alpha_{+\frac {1} {2} 0} |{+\frac {1} {2}}
  {0}\rangle   
  + \alpha_{+\frac {1} {2} -1} |{+\frac {1} {2}}
  {-1}\rangle  \cr
  &+& \alpha_{-\frac {1} {2} +1} |{-\frac {1} {2}}
  {+1}\rangle
  + \alpha_{-\frac {1} {2} 0} |{-\frac {1} {2}}
  {0}\rangle  
  + \alpha_{-\frac {1} {2} -1} |{-\frac {1} {2}}
  {-1}\rangle  ~,
\label{eq13_a0}
\end{eqnarray}
with complex expansion coefficients $\alpha_{m_n m_d}$ listed in
Table~\ref{tab1}. They satisfy the normalization condition
$$\sum_{m_n m_d} |\alpha_{m_n m_d}|^2=1~.$$ 
Since these states are cross-product states, it is evident that they are
not entangled.

The complex coefficients $\alpha_{m_n m_d}$ corresponding to opposite values
of the neutron and deuteron spin projections are related by:
\begin{eqnarray}
 {\alpha}_{-\frac {1} {2} -1}  &=& 
 \mp i \ {\alpha}_{+\frac {1} {2} +1}  \cr
 {\alpha}_{-\frac {1} {2} +1}  &=& 
 \mp i \ {\alpha}_{+\frac {1} {2} -1} \cr
 {\alpha}_{-\frac {1} {2} 0}  &=& 
 \pm i \ {\alpha}_{+\frac {1} {2} 0}  ~,
\label{eq13_a0_cin}
\end{eqnarray}
with a minus sign for the incoming neutron–deuteron polarization
combinations a), b), and f) of Table~\ref{tab1}, and a plus sign
for combinations c), d), and e).

%%%%%%%%%%%%%%%%%%%%%%%%%%%%%%%%%%

\begin{table}[ht!]
\centering  
\begin{tabular}{ |c|c|c|c|c|c|c| }
\hline
$p_y^n$, $p_y^d$, and  $p_{yy}$ combinations & $\alpha_{+\frac {1} {2} +1  }$
& $\alpha_{+\frac {1} {2} 0 } $  & $\alpha_{+\frac {1} {2} -1  } $ 
& $\alpha_{-\frac {1} {2} +1 }$ & $\alpha_{-\frac {1} {2} 0 }$ &
$\alpha_{-\frac {1} {2} -1 }$ \\
\hline
a) $p_y^n=+1~p_y^d=+1~p_{yy}=+1$  & $-\frac {i\sqrt{2}} {4}$
& $+\frac {1} {2} $
& $+\frac {i\sqrt{2}} {4} $  & $+\frac {\sqrt{2}} {4}$ & $+\frac {i} {2}$ 
& $-\frac {\sqrt{2}} {4}$ \\
\hline
b) $p_y^n=+1~p_y^d=-1~p_{yy}=+1$  & $-\frac {i\sqrt{2}} {4}$
& $-\frac {1} {2}  $
&  $+\frac {i\sqrt{2}} {4} $  & $+\frac {\sqrt{2}} {4}$  & $-\frac {i} {2}$
& $-\frac {\sqrt{2}} {4}$ \\
\hline
c) $p_y^n=-1~p_y^d=+1~p_{yy}=+1$  & $+\frac {i\sqrt{2}} {4}$
&  $-\frac {1} {2} $ 
&  $-\frac {i\sqrt{2}} {4} $    & $+\frac {\sqrt{2}} {4}$   & $+\frac {i} {2}$
& $-\frac {\sqrt{2}} {4}$\\
\hline
d) $p_y^n=-1~p_y^d=-1~p_{yy}=+1$  & $+\frac {i\sqrt{2}} {4}$
& $+\frac {1} {2}  $ 
& $-\frac {i\sqrt{2}} {4}$   & $+\frac {\sqrt{2}} {4}$   & $-\frac {i} {2}$
& $-\frac {\sqrt{2}} {4}$ \\
\hline
e) $p_y^n=+1~p_y^d=~~0~p_{yy}=-2$  & $+\frac {i\sqrt{2}} {4}$
&  $0$
&  $+\frac {i\sqrt{2}} {4} $  &  $-\frac {\sqrt{2}} {4}$  & $0$
&   $-\frac {\sqrt{2}} {4}$ \\
\hline
f) $p_y^n=-1~p_y^d=~~0~p_{yy}=-2$  &  $-\frac {i\sqrt{2}} {4}$
& $0$
& $-\frac {i\sqrt{2}} {4} $   & $-\frac {\sqrt{2}} {4}$  & $0$
&  $-\frac {\sqrt{2}} {4}$ \\
\hline
\end{tabular}
\caption{Values of the complex coefficients $\alpha_{m_n m_d}$ for the six
  pure initial states (\ref{eq13_a0}), corresponding to different combinations
  of the neutron and deuteron polarizations.}
\label{tab1}
\end{table}

\section{Results and discussion}
\label{results}

To investigate the pure final spin states in $nd$ elastic scattering
and the deuteron breakup reaction, formed from the six pure initial states
of Eq.~(\ref{eq13_a0}) with complex expansion coefficients $\alpha_{m_n m_d}$
listed in Table~\ref{tab1}, we solved the 3N Faddeev equation~(\ref{eq1a})
at four incoming neutron laboratory energies: $E = 8.5$, $37.5$, $65$,
and $135$~MeV. We neglected the three-nucleon force (3NF) in the 3N
nuclear Hamiltonian and restricted ourselves to pairwise interactions
between nucleons, taken as the CD Bonn NN potential~\cite{cdb}. Based
on these solutions, the elastic scattering and half-on-the-energy-shell
breakup amplitudes were obtained, which are required for the calculation
of the final spin density matrices in both reactions.

\subsection{Elastic nd scattering}
\label{elas}

The pure final state generated in $nd$ elastic scattering from the pure
initial state $|\psi_{nd}^{in,spin}\rangle$ is, according to
Eq.~(\ref{new6}), given by:
\begin{eqnarray}
  \   |\psi^{nd} \rangle &=& 
  \sum_{m_nm_d} \frac {\langle \frac {1} {2} m_n 1 m_d | T
    |\psi_{nd}^{in,spin} \rangle }
      {[\text{Tr} (T  |\psi_{nd}^{in,spin} \rangle
      \langle \psi_{nd}^{in,spin} | T^{\dagger} )]^{\frac {1} {2} }} ~
      | \frac {1} {2} m_n 1 m_d \rangle
 = \sum_{m_nm_d} {\bar{\alpha}}_{m_nm_d} ~
  | \frac {1} {2} m_n 1 m_d \rangle  ~,
\label{new_a1} 
\end{eqnarray}
where 
\begin{eqnarray}
{\bar{\alpha}}_{m_nm_d} &\equiv&
 \sum_{m_n'm_d'} {{\alpha}}_{m_n'm_d'} \frac {\langle \frac {1} {2} m_n 1 m_d | T
    |m_n'm_d' \rangle }
      {[\text{Tr} (T  |\psi_{nd}^{in,spin} \rangle
      \langle \psi_{nd}^{in,spin} | T^{\dagger} )]^{\frac {1} {2} }}  ~.
\label{new_a1_1} 
\end{eqnarray}

Among all pure $nd$ final spin states generated at different angles from
the six pure initial states listed in Table~\ref{tab1}, we searched for
entangled ones—those resembling the maximally entangled Bell states of
two-qubit systems~\cite{bookqinf}.
By entanglement we mean that the outcomes of neutron and deuteron
spin-projection measurements in such a state are strongly correlated: the
result of the deuteron spin projection measurement is uniquely determined
by the outcome of the neutron spin projection measurement.

The condition for identifying such a state is that, among the three available
pairs of complex expansion coefficients
$(\bar{\alpha}_{m_n m_d},\bar{\alpha}_{-m_n -m_d})$ with opposite
neutron and deuteron spin projections, only one pair contributes
significantly in Eq.~(\ref{new_a1}).
For the $nd$ system, two such Bell-like states are possible—one based
on the pair $(\bar{\alpha}_{+\frac{1}{2}, +1}, \bar{\alpha}_{-\frac{1}{2}, -1})$:
\begin{eqnarray}
  |\psi_{I}^{nd} \rangle &=& \bar{\alpha}_{+\frac {1} {2} +1} (~
  | \frac {1} {2} , { +\frac {1} {2}} , 1 , {+1}\rangle
  \mp i~ | \frac {1} {2} , {-\frac {1} {2}} , 1 , {-1} \rangle~ )  ~,
\label{eq13_a0_a}
\end{eqnarray}
and second on 
$(\bar{\alpha}_{+\frac {1} {2} -1  },\bar{\alpha}_{-\frac {1} {2} +1  })$:
\begin{eqnarray}
  |\psi_{II}^{nd} \rangle &=& \bar{\alpha}_{+\frac {1} {2} -1} (~
  | \frac {1} {2} , { +\frac {1} {2}} , 1 , {-1}\rangle
  \mp i~ | \frac {1} {2} , {-\frac {1} {2}} , 1 , {+1} \rangle~ )  ~,
\label{eq13_a0_b}
\end{eqnarray}
with a minus sign for final states resulting from incoming neutron–deuteron
polarization combinations a), b), and f) of Table~\ref{tab1}, and a
plus sign for c), d), and e).
In (\ref{eq13_a0_a}) and (\ref{eq13_a0_b}), we applied the relations between
$\bar{\alpha}$ coefficients with opposite spin projections:
\begin{eqnarray}
 \bar{\alpha}_{-\frac {1} {2} -1}  &=& 
 \mp i \ \bar{\alpha}_{+\frac {1} {2} +1}  \cr
 \bar{\alpha}_{-\frac {1} {2} +1}  &=& 
 \mp i \ \bar{\alpha}_{+\frac {1} {2} -1}  ~.
\label{eq13_a0_c}
\end{eqnarray}
This is valid, with the same comment regarding the $\mp$ sign.
They follow from (\ref{eq13_a0_cin}) and from the relation between
$nd$ elastic scattering
transition matrix elements with opposite neutron and deuteron
spin projections:
\begin{eqnarray}
{\langle \frac {1} {2} -m_n 1 -m_d | T
  |-m_n'-m_d' \rangle }&=& (-1)^{m_n'+m_d'-m_n-m_d} 
{\langle \frac {1} {2} m_n 1 m_d | T
    |m_n'm_d' \rangle } ~
\label{eq13_a0_c_tr_el}
\end{eqnarray}

It is evident that the state based on the pair
$(\bar{\alpha}_{+\frac {1} {2} 0  },\bar{\alpha}_{-\frac {1} {2} 0  })$,
\begin{eqnarray}
  |\psi_{III}^{nd} \rangle &=& \bar{\alpha}_{+\frac {1} {2} 0} (~
  | \frac {1} {2} , { +\frac {1} {2}} , 1 , {0}\rangle
  \pm i~ | \frac {1} {2} , {-\frac {1} {2}} , 1 , {0} \rangle~ )  ~,
\label{eq13_a0_d}
\end{eqnarray}
which also contributes to the outgoing states, is not entangled.
Here we use the following relation:
\begin{eqnarray}
 \bar{\alpha}_{-\frac {1} {2} 0}  &=& 
 \pm i \ \bar{\alpha}_{+\frac {1} {2} 0}  ~.
\label{eq13_a0_e}
\end{eqnarray}

It follows that the condition for the existence of an entangled final $nd$
state requires, at a particular c.m. angle, that the contributions of
the state (\ref{eq13_a0_d}) and either the state (\ref{eq13_a0_a})
or (\ref{eq13_a0_b}) be negligible when compared to the dominant component,
given by either the state (\ref{eq13_a0_b}) or (\ref{eq13_a0_a}).

At this point, we need to justify our choice of pairs
$(\bar{\alpha}_{m_n m_d},\bar{\alpha}_{-m_n -m_d})$ used in the definition
of entangled contributions to the final state (\ref{new_a1}).
The deuteron has three possible spin projections, and our convention—to
treat the deuteron as a binary system by singling out the two states with
spin projections $m_d = \pm 1$—is reasonable only if the non-entangled
component (\ref{eq13_a0_d}), which in our framework represents
the entanglement-breaking part, vanishes.

In fact, one could instead select any two of the three deuteron
spin states with different spin projections as the effective binary system.
This would lead to two additional sets of pairs,
$(\bar{\alpha}_{m_n m_d},\bar{\alpha}_{-m_n {m'}_d})$, with ${m'}_d \ne m_d$,
producing four other classes of entangled states. The corresponding
entanglement-breaking components (\ref{eq13_a0_d}) would then be
constructed from the pairs of states
$$( | \frac {1} {2} , { +\frac {1} {2}} , 1 , {\pm 1} \rangle>,
 | \frac {1} {2} , { -\frac {1} {2}} , 1 , {\pm 1} \rangle>),$$ 
with the upper or lower sign corresponding to each additional set. Requiring
the vanishing of this entanglement-breaking component would,
however, also eliminate entanglement in the other two contributions.

\begin{figure}
\includegraphics[scale=0.65]{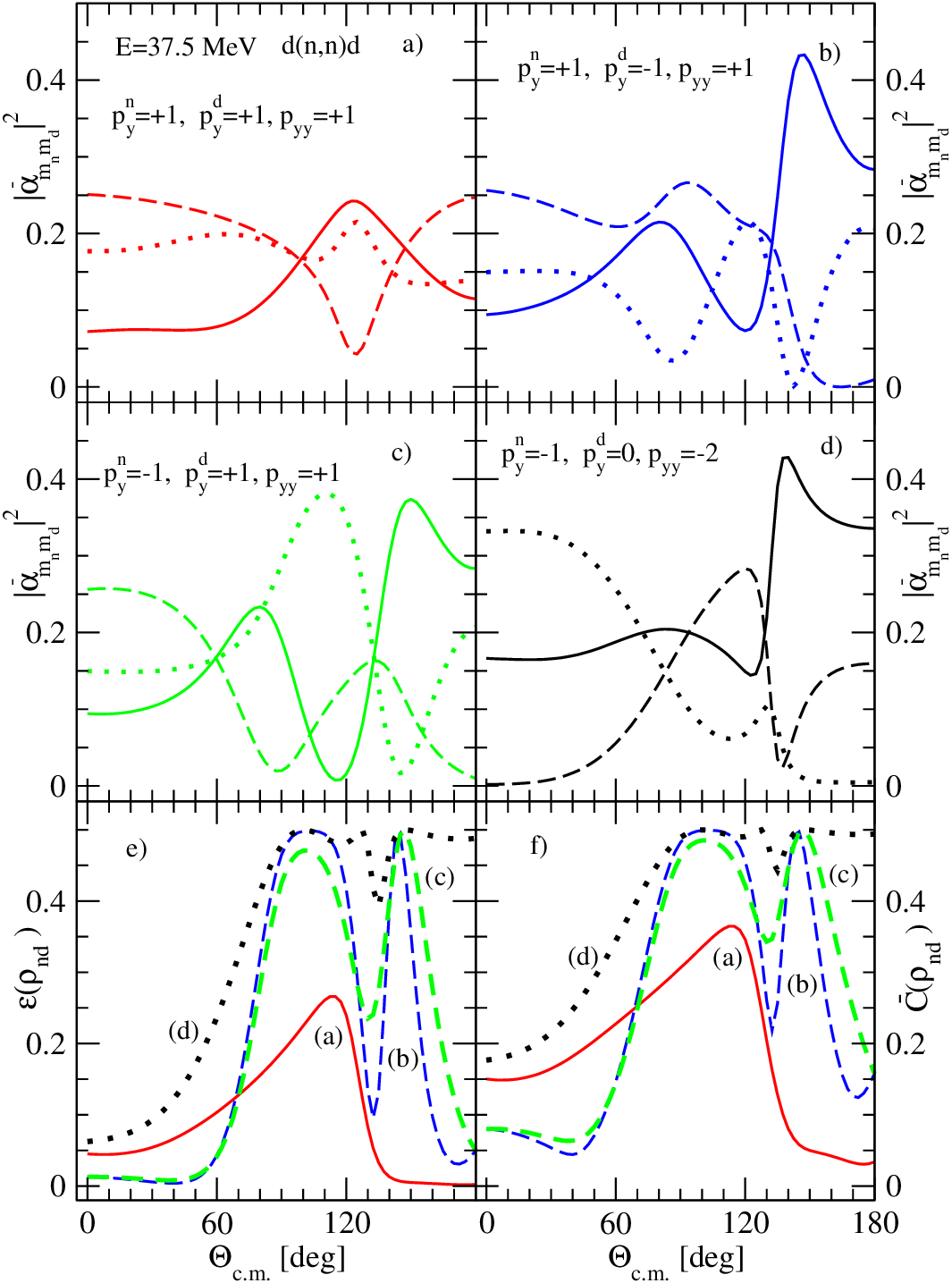}  
\caption{
  (color online)
  Angular distributions of the contributions $|\bar{\alpha}_{m_n m_d}|^2$ to
  the norm for the pure final $nd$ states of Eq.~(\ref{new_a1}).
  These states are formed in the $\vec d (\vec n, n)d$ scattering at
  an incoming neutron laboratory energy of $E = 37.5$~MeV, from pure
  cross-product initial states with the neutron and deuteron polarizations 
 $(p_y^n,p_y^d,p_{yy})$ shown in a)-d). 
The solid, dashed, and dotted lines in a)–d) show the contributions of
$|\bar{\alpha}_{+\frac{1}{2} +1}|^2 = |\bar{\alpha}_{-\frac{1}{2} -1}|^2$,
$|\bar{\alpha}_{+\frac{1}{2} 0}|^2 = |\bar{\alpha}_{-\frac{1}{2} 0}|^2$,
and $|\bar{\alpha}_{+\frac{1}{2} -1}|^2 = |\bar{\alpha}_{-\frac{1}{2} +1}|^2$,
respectively.
The angular distributions of the entanglement power $\epsilon(\rho_{nd})$
and the concurrence ${\bar C}(\rho_{nd})$ for a)–d) are presented in e)
and f), by the solid, short-dashed, long-dashed, and dotted lines of
the same colors as in a), b), c), and d), respectively.
The predictions were obtained using the CD-Bonn potential.}
\label{fig1}
\end{figure}

In Figs.~\ref{fig1}a)–\ref{fig1}d) we present, for $E = 37.5$~MeV, the
angular distributions of the contributions from each of the three components,
$$|\bar{\alpha}_{+\frac {1} {2} +1 }|^2=|\bar{\alpha}_{-\frac {1} {2} -1 }|^2,~~
 |\bar{\alpha}_{+\frac {1} {2} 0 }|^2=|\bar{\alpha}_{-\frac {1} {2} 0 }|^2,~~
    |\bar{\alpha}_{+\frac {1} {2} -1 }|^2=|\bar{\alpha}_{-\frac {1} {2} +1 }|^2,$$
to the norm of the final states. These contributions vary with the
scattering angle, and their relative magnitudes change depending on
the polarization of the incoming state.

In the case shown in Fig.~\ref{fig1}a), where all three contributions
are of comparable magnitude, it appears rather unlikely that an
entangled state could be found at any angle. Figure~\ref{fig1}a)
corresponds to combination a) of the initial neutron and deuteron
polarizations from Table~\ref{tab1}.

In contrast, Figs.~\ref{fig1}b), \ref{fig1}c), and \ref{fig1}d),
corresponding to polarization combinations b), c), and f)
from Table~\ref{tab1}, respectively, exhibit conditions under
which entangled states are likely to occur at backward angles,
where a clear maximum of a single component indicates its dominance.

Restricting the analysis solely to identifying the locations of
dominant contributions from a single component might be too limiting.
Therefore, we also examined whether quantifying the degree of spin
entanglement in the outgoing $nd$ pair—using two entanglement
measures proposed in the literature for the $np$ system in
neutron–proton scattering—would lead to similar locations of
entangled states as those obtained from the analysis
based on the magnitudes of the contributions to the norm.

To this end, we employed the entanglement power $\epsilon$, defined
in Refs.~\cite{beane2019,bai2022,bai2024,kirch2,liu2023}, and
the concurrence $C$, introduced in Refs.~\cite{liu2023,bai2024}.
These measures quantify the degree of entanglement for states that
are neither maximally entangled nor simple product states.

For the reader’s convenience, we briefly recall the relevant definitions.
The spin state of a two-particle system—such as the outgoing $nd$ pair
in $nd$ scattering—is generally described by the density
matrix $\rho_{nd}$, which has unit trace.
A quantity that characterizes the degree of entanglement of this state
is the entanglement power, $\epsilon(\rho_{nd})$, defined as
\begin{eqnarray}
\epsilon(\rho_{nd}) &=& 1 - \text{Tr}(\rho_n^2) ~,
\label{a_eq1}
\end{eqnarray}
where $\rho_{n} \equiv \text{Tr}_{d}(\rho_{nd})$ is the reduced density
matrix obtained by tracing $\rho_{nd}$ over the deuteron subsystem $d$.

Another measure used to quantify the strength of entanglement is
the concurrence, $\bar{C}(\rho_{nd})$, as proposed in Ref.~\cite{bai2024}:
\begin{eqnarray}
  {\bar C}(\rho_{nd}) &\equiv& \frac {1} {2}~
  \sqrt {1 - \langle \hat{\sigma}_x^n \rangle^2
    - \langle \hat{\sigma}_y^n \rangle^2
    - \langle \hat{\sigma}_y^n \rangle^2 }
  =   \frac {1} {2}~
  \sqrt {1  - \langle \hat{\sigma}_y^n \rangle^2 }   ~,
  \label{a_eq2a}
\end{eqnarray}
with the polarizations of the outgoing neutrons,
$\langle \hat{\sigma}_i^n \rangle$.
In coplanar reactions, due to parity conservation, only the component
$\langle \hat{\sigma}_y^n \rangle$ does not vanish.

These measures reach their maximum value of $0.5$ for a maximally
entangled state.
In addition, the entanglement power vanishes for a product spin state, and
the same holds for the concurrence.
Since the density matrix $\rho_{nd}$ depends on the c.m.
angle $\Theta_{c.m.}$, the quantities $\epsilon(\rho_{nd})$
and $\bar{C}(\rho_{nd})$ are also angle-dependent.

In Figs.~\ref{fig1}e) and \ref{fig1}f), we present the angular distributions
of the
entanglement power $\epsilon(\rho_{nd})$ and the concurrence $\bar{C}(\rho_{nd})$
at an energy of $E = 37.5$~MeV for the final pure $nd$ states shown in
Figs.~\ref{fig1}a)–\ref{fig1}d), corresponding to the four initial
neutron and deuteron
polarization combinations a), b), c), and f) listed
in Table~\ref{tab1}, respectively.

Up to about $\Theta_{c.m.} \approx 80^\circ$, both the entanglement power
and the concurrence remain small for all polarization combinations,
indicating that, in the forward-angle region at this energy, no entangled
states are expected to occur.
For the states corresponding to Fig.~\ref{fig1}a), the entanglement power
and concurrence also remain small at larger angles,
which supports the result obtained from the analysis of the magnitudes
of the contributions to the norm—
indeed, no entangled state was found for polarization combination a)
at this energy.

For combinations b), c), and f) (where f) corresponds to the states
in Fig.~\ref{fig1}d)),
the analysis of the magnitudes of the contributions to the norm yielded
one (state No.~3 in Table~\ref{tab2}),
two (state No.~4 and another at $\Theta_{c.m.} = 110^\circ$,
not listed in Table~\ref{tab2} due to a large,
$\approx 40 \%$, entanglement-breaking contribution),
and one (state No.~5) entangled states, respectively.
Their number and angular positions are consistent with the number
and locations of the corresponding maxima
of the entanglement power observed in Fig.~\ref{fig1}e).

It should be emphasized, however, that the two additional maxima of
the entanglement power visible in Fig.~\ref{fig1}e)
for polarization combinations (b) and (d), which reach values close to
 $0.5$ at $\Theta_{c.m.} \approx 110^\circ$,
do not correspond to genuine entangled states.
The pure final states around this angle, for these two particular
polarization combinations,
are dominated by contributions from $|\psi_{I}^{nd}\rangle$ and, to
an even greater extent,
from the entanglement-breaking component (\ref{eq13_a0_d}).
This clearly illustrates that a reliable identification of entanglement
requires considering
all available information about the state, rather than relying on a
single diagnostic measure.

The assertion that an entangled state exists at a particular angle implies
that such states are also present within a certain region around that angle,
whose extent depends on both the energy and the type of state considered.
The use of the entanglement power offers a twofold advantage: in addition
to identifying the location of an entangled state, it also provides
information about the range of angles over which the states remain entangled.
The angular widths of the maxima of the entanglement power, at which the
entangled states occur, provide a quantitative measure of this range.

It is also interesting to note the strong similarity and clear correlation
between the angular dependencies of the entanglement power and the concurrence.
This consistency ensures that analyses based on either quantity lead to
the same conclusions regarding both the number and the locations of
the entangled states.

In Table~\ref{tab2}, we list all entangled states identified among the
final pure spin $nd$ states generated from six combinations of polarizations
of pure incoming states at the four considered energies.
The analysis was performed by examining both the magnitudes of
the contributing components and the values of the entanglement power.

We assumed that a significant contribution to the norm from a particular
pair of amplitudes—either
$(\bar{\alpha}_{+\frac {1}{2} +1}, \bar{\alpha}_{-\frac {1}{2} -1})$ or
$(\bar{\alpha}_{+\frac {1}{2} -1}, \bar{\alpha}_{-\frac {1}{2} +1})$—
must exceed approximately $70 \%$, that is,
$$|\bar{\alpha}_{m_n m_d}|^2 + |\bar{\alpha}_{-m_n -m_d}|^2 \gtrsim 0.7 ~.$$
The angle $\Theta_{c.m.}$ shown in Table~\ref{tab2} corresponds to the position
where a maximum of the entanglement power occurs and where the maximal
contribution from the relevant pair of amplitudes is found.

With this threshold value of $0.7$, we identified twelve entangled
Bell-like states of both types,
$|\psi_{I}^{nd} \rangle$ and $|\psi_{II}^{nd} \rangle$.
Some of these states exhibit relatively large contributions from
the entanglement-spoiling components, reaching approximately $30 \%$
of the norm.
Reducing the limiting value of the entanglement-violating component
from $0.3$ to $0.2$ significantly decreases the number of entangled
states—from twelve to six.
Further lowering it to $0.1$ leaves only a single state, No.~8, at
$E = 135$~MeV.
This indicates that it is extremely difficult, if not impossible,
to identify final $nd$ states that remain entangled while having only
a very small contribution from the entanglement-spoiling component.

\begin{figure}
\includegraphics[scale=0.65]{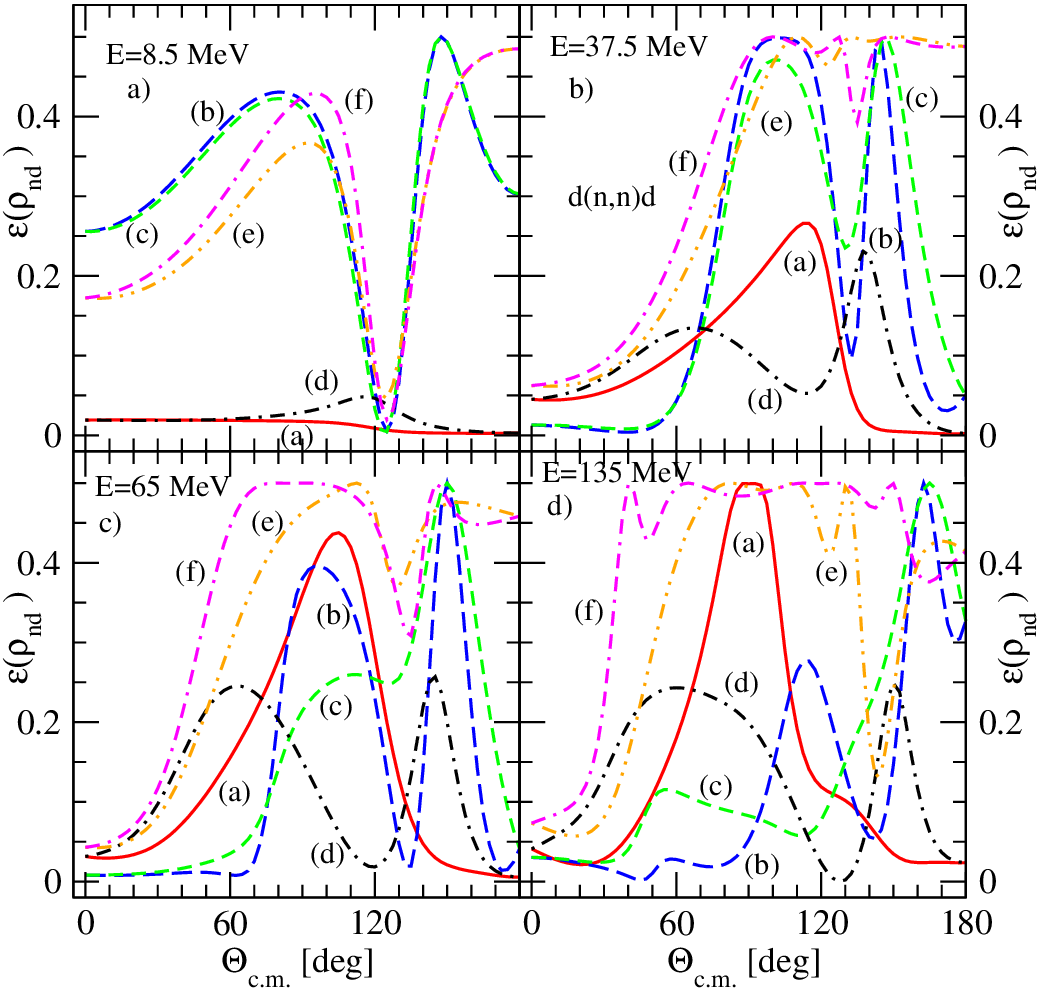}  
\caption{
  (color online)
  Angular distributions of the entanglement power $\epsilon(\rho_{nd})$ for
  the elastic scattering pure final spin states of Eq.~(\ref{new_a1}).
  These states are formed in the $\vec d (\vec n, n)d$ scattering at
  incoming neutron laboratory energies of $E = 8.5$~MeV (a),
  $E = 37.5$~MeV (b), $E = 65$~MeV (c), and $E = 135$~MeV (d),
  from pure cross-product initial states with the neutron and deuteron
  polarization combinations given in Table~\ref{tab1}.
  Their entanglement power $\epsilon(\rho_{nd})$ is shown by different
  lines: (a) red solid, (b) blue long-dashed, (c) green short-dashed,
  (d) black dash-dotted, (e) orange double-dash-dotted,
  and (f) magenta double-dash-dotted.
The predictions were obtained using the CD-Bonn potential.}
\label{fig1a}
\end{figure}

In Figs.~\ref{fig1a}a)–\ref{fig1a}d), we present the angular distributions
of the entanglement power for all initial polarization combinations
at all studied energies.
At $E = 8.5$~MeV (Fig.~\ref{fig1a}a)), the small values of the entanglement
power exclude the possibility of finding entangled states for the initial
polarization combinations a) and d). Only combinations b) and c) yield
one entangled state each, located at the angles listed in Table~\ref{tab2}.

Considering again the energy $E = 37.5$~MeV and the polarization
combinations not shown in Fig.~\ref{fig1}e), one might wonder whether
combination e) (orange dash–double-dotted line in Fig.~\ref{fig1a}b))
was accidentally omitted from Table~\ref{tab2}.
However, an examination of the contributions to the norm around
$\Theta_{c.m.} \approx 120^\circ$ shows that the dominant component
at this angle contributes less than $0.7$, and therefore does not meet
the adopted threshold.

At $E = 65$~MeV, the behavior of the entanglement power supports
the positions of the two entangled states listed in Table~\ref{tab2}:
one for polarization combination e) at $112^\circ$ and another for
combination f) at $146^\circ$.
For combination f), the entanglement power reaches an even larger
value around $115^\circ$, yet no entangled state exists at this angle,
since all three components contribute comparably to the norm.
Similarly, the pronounced maxima of the entanglement power around
$150^\circ$ for combinations b) and c), which even exceed
the corresponding value for the entangled state No.~7 (combination f)),
require clarification.
In both cases, the contribution of the leading component reaches only $0.6$,
which is below the $0.7$ threshold adopted for the states listed
in Table~\ref{tab2}.

These cases demonstrate that relying solely on the values of
the entanglement power may be misleading; both the behavior of
the entanglement power and the magnitudes of the components contributing
to the norm must be considered when identifying entangled states.

At $E = 135$~MeV, entangled states were found for polarization
combinations a), e), and f).
The small values of the entanglement power for combination d) exclude
the existence of entangled states for this configuration.
Similarly, for b) and c), no entangled states are observed below
$\Theta_{c.m.} \approx 150^\circ$.
The large maxima at very backward angles (around $160^\circ$) for
combinations b) and c) correspond to states dominated by
the non-entangled component $|\psi_{III}^{nd}\rangle$.

Among the states listed in Table~\ref{tab2}, state No.~8 at $E = 135$~MeV
has the smallest contribution—about $8.8 \%$ of the norm—from the
entanglement-spoiling components.
Is such a contribution small enough to consider this state as purely
entangled?
To address this question, we calculated the expectation values of
selected spin observables in our states and compared them with
the corresponding values for a perfectly entangled state.

The following observables were considered:
the neutron polarization,
$$\langle \sigma_y^n \rangle = \langle \psi^{nd} | \sigma_y^n \otimes I^d
| \psi^{nd}  \rangle,$$
the neutron–deuteron spin correlations,
$$\langle \sigma_y^n \otimes P_{yy} \rangle = \langle \psi^{nd} |
\sigma_y^n \otimes P_{yy} | \psi^{nd} \rangle,
\langle \sigma_y^n \otimes P_{y}^2 \rangle = \langle \psi^{nd} |
\sigma_y^n \otimes P_{y}^2 | \psi^{nd} \rangle,$$
and the deuteron polarization,
$$ \langle P_{y} \rangle = \langle \psi^{nd} |
I^n \otimes P_{y} | \psi^{nd} \rangle .$$ 

A direct calculation gives:
\begin{eqnarray}
\langle \sigma_y^n  \rangle &\equiv& \langle \psi^{nd} |
\sigma_y^n \otimes I^d | \psi^{nd} \rangle
   \cr
&=& |\bar{\alpha}_{+\frac {1} {2} 0 }|^2 [ \pm 2 ] +
2 Re \{ \bar{\alpha}_{+\frac {1} {2} +1 }^* \bar{\alpha}_{+\frac {1} {2} -1 }
[\mp 2 ] \}  ~,
\label{eq_cor1}
\end{eqnarray}
\begin{eqnarray}
\langle \sigma_y^n \otimes P_{yy} (P_y^2) \rangle &\equiv& \langle \psi^{nd} |
\sigma_y^n \otimes P_{yy} (P_y^2) | \psi^{nd} \rangle
=|\bar{\alpha}_{+\frac {1} {2} +1 }|^2 [\pm 3 (\pm 1) ] + 
|\bar{\alpha}_{+\frac {1} {2} -1 }|^2 [\pm 3 (\pm 1) ]  \cr
&+&
|\bar{\alpha}_{+\frac {1} {2} 0 }|^2 [\pm 2  (\pm 2) ] +
2 Re \{ \bar{\alpha}_{+\frac {1} {2} +1 }^* \bar{\alpha}_{+\frac {1} {2} -1 }
[\pm 1 (\mp 1) ] \} \cr
&=& \pm \frac {3} {2} (\pm \frac {1} {2} ) + 
|\bar{\alpha}_{+\frac {1} {2} 0 x}|^2 [\mp 1  (\pm 1) ] +
2 Re \{ \bar{\alpha}_{+\frac {1} {2} +1 }^* \bar{\alpha}_{+\frac {1} {2} -1 }
[\pm 1 (\mp 1) ] \}  ~,
\label{eq_cor2}
\end{eqnarray}
and
\begin{eqnarray}
\langle  P_y \rangle &\equiv& \langle \psi^{nd} |
 I^n \otimes P_y | \psi^{nd} \rangle
   \cr
   &=& 2 \sqrt {2} Im [ (\bar{\alpha}_{+\frac {1} {2} +1 }^* -
   \bar{\alpha}_{+\frac {1} {2} -1 }^* ) \bar{\alpha}_{+\frac {1} {2} 0 } ]  ~.
\label{eq_cor3}
\end{eqnarray}

Since either $\bar{\alpha}_{+\frac{1}{2} +1}$ or $\bar{\alpha}_{+\frac{1}{2} -1}$,
together with $\bar{\alpha}_{+\frac{1}{2} 0}$, are entanglement-spoiling
components, they both vanish for purely entangled states.
Consequently, the neutron and deuteron polarizations in such states also
vanish, while the spin correlations take the values
$$\langle \sigma_y^n \otimes P_{yy}  \rangle = \pm \frac {3} {2}, ~~~
\langle \sigma_y^n \otimes P_y^2 \rangle = \pm \frac {1} {2}.$$

It is interesting to note that deviations from these ideal values in our
states are proportional to the product of the amplitudes $\alpha$
corresponding to the dominant and entanglement-spoiling components.

In the last four columns of Table~\ref{tab2}, we show the actual values
of the spin correlations and polarizations for all states.
While the spin correlations and the neutron polarization exhibit only
small deviations from the values expected for purely entangled states,
the deuteron polarization differs significantly from zero, taking values
between approximately $0.1$ and $0.5$.
Even for state No.~8, which has the smallest contribution from
the entanglement-spoiling components, $\langle P_y \rangle = 0.147$,
the deviation from zero remains noticeable.
As seen in Table~\ref{tab2}, this behavior results from the relatively
large contribution of the entanglement-breaking
amplitudes $\bar{\alpha}_{\pm \frac{1}{2} 0}$.
This indicates that the states identified in $nd$ elastic scattering
cannot be regarded as highly entangled pure states.

\subsection{Exclusive breakup} 
\label{breakup}

The final pure spin state of the three free outgoing nucleons, created
from six pure initial states in the exclusive $nd$ breakup reaction
$\vec{d}(\vec{n}, N_1 N_2) N_3$, is given, according to (\ref{new6}), by:
\begin{eqnarray}
  \   |\psi^{br} \rangle  &=& 
  \sum_{m_1m_2m_3} \frac {\langle \frac {1} {2} m_1 \frac {1} {2} m_2
    \frac {1} {2} m_3 | T  | \psi_{nd}^{in,spin} \rangle }
      {[\text{Tr} (T  | \psi_{nd}^{in,spin} \rangle 
      \langle \psi_{nd}^{in,spin} | T^{\dagger} )]^{\frac {1} {2} }} ~
      | \frac {1} {2} m_1 \frac {1} {2} m_2 \frac {1} {2} m_3 \rangle \cr
   &\equiv& \sum_{m_1m_2m_3} {\bar{\alpha}}_{m_1m_2m_3} ~
  | \frac {1} {2} m_1 \frac {1} {2} m_2 \frac {1} {2} m_3 \rangle  ~, 
\label{new_a2} 
\end{eqnarray}
with eight complex coefficients ${\bar{\alpha}}_{m_1 m_2 m_3}$, using a notation
in which, in the following, we will denote the nucleon spin projections
not by their values $\pm \frac{1}{2}$, but simply by $\pm$.

The exclusive breakup reaction provides infinitely many kinematically
complete configurations of the three outgoing nucleons.
In practice, a particular configuration is selected by placing two nucleon
detectors at specific laboratory angles
$(\theta_1^{\rm lab}, \phi_1^{\rm lab})$ and $(\theta_2^{\rm lab}, \phi_2^{\rm lab})$,
and by measuring the energies of nucleons $N_1$ and $N_2$ in coincidence.
Conservation of energy and momentum ensures that the experimental events
then lie in the energy plane $E_1^{\rm lab}$–$E_2^{\rm lab}$ of nucleons
$N_1$ and $N_2$ along the so-called kinematically allowed S-curve.
The arclength $S$ along this curve, after choosing a reference point $S=0$
by some convention, parametrizes the position on the line.
At each position $S$, the momenta of the three final nucleons are
unambiguously determined~\cite{glo96}.

In this study, we focused on two coplanar, kinematically complete geometries
with dominant unpolarized cross sections.
The first is the so-called final-state-interaction (FSI) geometry, in which
the two outgoing nucleons, $N_1$ and $N_2$, have equal momenta~\cite{glo96}.
Their strong interaction in the $^1S_0$ partial-wave state leads, when viewed
along the S-curve, to a characteristic cross-section maximum occurring
under the exact FSI condition.
The magnitude of this maximum is sensitive to the $^1S_0$ scattering length.

The second is the quasi-free-scattering (QFS) geometry, which corresponds
to a situation in which nucleon $N_3$ remains at rest in the laboratory
frame, and interactions occur predominantly between nucleons $N_1$
and $N_2$~\cite{glo96}.
This leads, when viewed along the S-curve, to a characteristic cross-section
maximum at the exact QFS condition, the shape of which reflects
the distribution of relative momenta of nucleons
in the deuteron~\cite{glo96,wit1989}.

For the $nd$ breakup, the neutron–proton ($np$) or neutron–neutron ($nn$)
pairs can either interact strongly in the FSI geometry or emerge
as quasi-freely scattered pairs.

In order to investigate all QFS and FSI configurations, for both $np$ and $nn$
systems, we considered their dependence on the laboratory angle of the
first nucleon, $N_1$, denoted $\theta_1^{\rm lab}$, assuming that at this
angle the FSI and QFS conditions are exactly fulfilled (which, when
viewed along the S-curve, corresponds to the location where
the cross-section maximum occurs).

At the energies considered, some values of $\theta_1^{\rm lab}$ correspond
to two solutions that satisfy the FSI and QFS conditions.
The second solutions, which are difficult to access experimentally, form
a branch that does not extend to small values of $\theta_1^{\rm lab}$.
This branch corresponds, in the FSI case, to small energies of
the interacting nucleons, and, in the QFS case, to small values of
the angle $\theta_2^{\rm lab}$ and energy $E_1^{\rm lab}$.

Among all the final pure spin states for the FSI and QFS geometries, generated
from the six pure initial states listed in Table~\ref{tab1}, we searched for
entangled states that resemble maximally entangled Bell states in two-qubit
systems~\cite{bookqinf}.
By entanglement, we again mean that the results of the spin-projection
measurements in a chosen two-nucleon subsystem—here taken as the subsystem
$N_1$–$N_2$—must be strongly correlated: the outcome of the spin-projection
measurement for the second nucleon, $N_2$, must be unequivocally determined
by the measurement result of the first nucleon, $N_1$.

The condition for identifying such a state is that only two pairs, out of
the four available pairs of complex expansion coefficients,
 $$(\bar{\alpha}_{m_1 m_2 m_3},\bar{\alpha}_{-m_1 -m_2 -m_3}),$$ 
with opposite nucleon spin projections, contribute significantly
in (\ref{new_a2}).

To examine in more detail the form of the spin state in Eq.~(\ref{new_a2})
under the FSI(12) and QFS(12) conditions, we applied the following relations
between the amplitudes $\bar{\alpha}_{m_1 m_2 m_3}$ with opposite spin projections:
\begin{eqnarray}
 \bar{\alpha}_{---}  &=& 
 \mp i \ \bar{\alpha}_{+++}  \cr
 \bar{\alpha}_{--+}  &=& 
 \pm i \ \bar{\alpha}_{++-}  \cr
 \bar{\alpha}_{-+-}  &=& 
 \pm i \ \bar{\alpha}_{+-+}  \cr
 \bar{\alpha}_{-++}  &=& 
 \mp i \ \bar{\alpha}_{+--}  ~.
\label{eq13_a1_d}
\end{eqnarray}
These relations are valid for coplanar geometries, with the upper sign
corresponding to final states resulting from incoming neutron–deuteron
polarization combinations a), b), and f) of Table~\ref{tab1}, and
the lower sign for combinations c), d), and e).
They can be derived in the same manner as relations (\ref{eq13_a0_c})
and (\ref{eq13_a0_e}) for elastic $nd$ scattering.
Using these relations, the coplanar breakup spin state
in (\ref{new_a2}) can be expressed as:
\begin{eqnarray}
  \   |\psi^{br} \rangle  &=& 
  | + \rangle_3 \ ( \ \bar{\alpha}_{+++} \ | ++ \rangle_{12} 
  \pm i \ \bar{\alpha}_{++-} \ | -- \rangle_{12} \ ) \cr
 &+&  | - \rangle_3 \ ( \ \bar{\alpha}_{++-} \ | ++ \rangle_{12} 
  \mp i \ \bar{\alpha}_{+++} \ | -- \rangle_{12} \ ) \cr
 &+&  | + \rangle_3 \ ( \ \bar{\alpha}_{+-+} \ | +- \rangle_{12} 
  \mp i \ \bar{\alpha}_{+--} \ | -+ \rangle_{12} \ ) \cr  
 &+&  | - \rangle_3 \ ( \ \bar{\alpha}_{+--} \ | +- \rangle_{12} 
  \pm i \ \bar{\alpha}_{+-+} \ | -+ \rangle_{12} \ )  ~. 
\label{new_a4} 
\end{eqnarray}

From (\ref{new_a4}), it is evident that the vanishing of the first two pairs
of amplitudes, $(\bar{\alpha}_{+++}, \bar{\alpha}_{---})$ and
$(\bar{\alpha}_{++-}, \bar{\alpha}_{--+})$, would result, in
the $N_1$–$N_2$ subsystem, in the Bell-like state $|\psi_{I}^{\rm br} \rangle$:
\begin{eqnarray}
  |\psi_{I}^{br} \rangle &=& ~~
  ( \ | + \rangle_3  \ \bar{\alpha}_{+-+} \ +
  | - \rangle_3 \  \ \bar{\alpha}_{+--} \ ) 
  \ | +- \rangle_{12}  \cr
  &\mp& i \ ( \ | + \rangle_3  \ \bar{\alpha}_{+--} \ -
  | - \rangle_3 \  \ \bar{\alpha}_{+-+} \ )
  \ | -+ \rangle_{12}  
  ~,
\label{new_a5}
\end{eqnarray}
while the vanishing of the last two pairs of amplitudes,
$(\bar{\alpha}_{+-+}, \bar{\alpha}_{-+-})$ and
$(\bar{\alpha}_{+--}, \bar{\alpha}_{-++})$, would lead to the Bell-like
state $|\psi_{II}^{\rm br} \rangle$:
\begin{eqnarray}
  |\psi_{II}^{br} \rangle &=& ~~
  ( \ | + \rangle_3  \ \bar{\alpha}_{+++} \ +
  | - \rangle_3 \  \ \bar{\alpha}_{++-} \ ) 
  \ | ++ \rangle_{12}  \cr
  &\pm& i \ ( \ | + \rangle_3  \ \bar{\alpha}_{++-} \ -
  | - \rangle_3 \  \ \bar{\alpha}_{+++} \ )
  \ | -- \rangle_{12}  
  ~.
\label{new_a6}
\end{eqnarray}

We will see that the first scenario occurs to a very good approximation
for FSI($np$) and QFS($nn$) configurations and is caused by the dominant
contribution, in these geometries, from the singlet $^1S_0$ component
of the NN interaction.
For all FSI($nn$) configurations, this scenario is fulfilled exactly, which
causes all FSI($nn$) final spin states to be maximally entangled.

In this case, in addition to $\bar{\alpha}_{++-} = \bar{\alpha}_{+++} = 0$, we
also have $\bar{\alpha}_{+-+} = \pm i \bar{\alpha}_{+--}$, so that
the Bell-like FSI($nn$) entangled state $|\psi_{I}^{\rm FSI(nn)} \rangle$
can be written as:
\begin{eqnarray}
  |\psi_{I}^{FSI(nn)} \rangle &=& ~~
   \bar{\alpha}_{+-+}    \ ( \ | + \rangle_3  \ \mp i \
  | - \rangle_3 \ )
  \ ( \  | +- \rangle_{12}  \
  -   \ | -+ \rangle_{12} \ )   ~.
\label{new_a7}
\end{eqnarray}

We remind the reader that, in Eqs.~(\ref{new_a5})–(\ref{new_a7}), the upper
and lower $\pm$ signs refer to the incoming polarization combinations
of Table~\ref{tab1}, namely a), b), f) and c), d), e), respectively.

\begin{figure}
\includegraphics[scale=0.65]{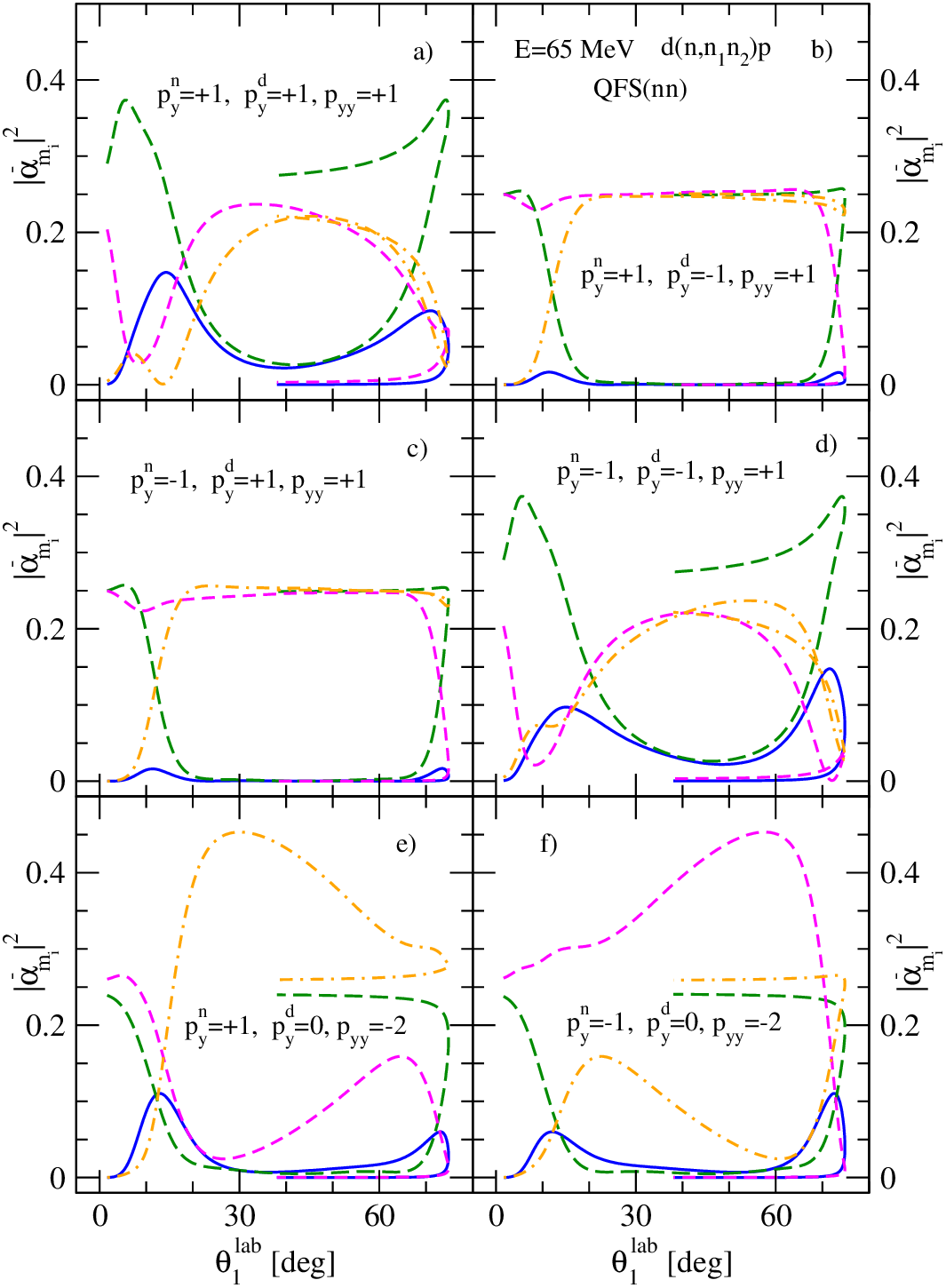}  
\caption{
  (color online)
  Contributions $|\bar{\alpha}_{m_1 m_2 m_3}|^2$ to the norm of the breakup
  pure final spin states of Eq.~(\ref{new_a4}) in QFS(nn) kinematically
  complete configurations, shown as functions of the laboratory angle
  of the first neutron $\theta_1^{lab}$.
  These states are formed in the $\vec d (\vec n, n_1 n_2)p$ breakup
  reaction at an incoming neutron laboratory energy of $E = 65$~MeV, from
  pure cross-product initial states with the neutron and deuteron
  maximal polarizations given in Table~\ref{tab1} and shown in a)-f).
  The blue solid, green long-dashed, magenta short-dashed, and orange
  dash-dotted lines in a)–f) show the contributions of
$|\bar{\alpha}_{---}|^2 = |\bar{\alpha}_{+++}|^2$,
$|\bar{\alpha}_{--+}|^2 = |\bar{\alpha}_{++-}|^2$,
$|\bar{\alpha}_{-+-}|^2 = |\bar{\alpha}_{+-+}|^2$,
and $|\bar{\alpha}_{-++}|^2 = |\bar{\alpha}_{+--}|^2$, respectively.
The predictions were obtained using the CD-Bonn potential.}
\label{fig2}
\end{figure}

In Figs.~\ref{fig2}a)–\ref{fig2}f), we illustrate, at $E=65$~MeV, the
contributions of the different components $|\bar{\alpha}_{m_1 m_2 m_3}|^2$
to the norm of the breakup pure final spin states in Eq.~(\ref{new_a2})
for all QFS($nn$) complete geometries.
At each angle $\theta_1^{\rm lab}$, the exact quasi-free-scattering condition,
$\vec{p}_p = 0$, is fulfilled.
The pure final states shown in Figs.~\ref{fig2}a)–\ref{fig2}f) are
generated from the initial pure states corresponding to neutron and
deuteron polarization combinations a)–f) of Table~\ref{tab1}.
In each case a)–f), the second branch of solutions for the given
$\theta_1^{\rm lab}$, which does not extend to smaller values of this angle,
is clearly visible.

The conspicuous approach to zero of the blue solid and green long-dashed
lines in Fig.~\ref{fig2} over wide regions of the angle $\theta_1^{\rm lab}$,
particularly in b) and c), indicates small values of the amplitudes
$\bar{\alpha}_{---}$, $\bar{\alpha}_{+++}$, $\bar{\alpha}_{--+}$,
and $\bar{\alpha}_{++-}$ at these locations.
Consequently, this implies that the Bell-like entangled states
$|\psi_{I}^{\rm br} \rangle$ of (\ref{new_a5}) must appear in the
QFS($nn$) complete geometries, at least at an energy of $E=65$~MeV.

The appearance of such states at any energy follows from the fact that
quasi-free nucleon–nucleon scattering in the deuteron breakup reaction
closely resembles free $NN$ scattering, with the similarity increasing
with the incoming neutron energy.
The smallness of the above amplitudes reflects the absence of
the spin-$s=1$ component in free $nn$ scattering, where only $s=0$ contributes.

Investigation of the magnitudes of the different contributions to
the QFS($nn$) wave function (\ref{new_a4}) at our four energies supports
the above conclusion. Indeed, we found a multitude of Bell-like entangled
states in the QFS($nn$) geometries over wide regions of the angle
$\theta_1^{\rm lab}$, all of type $|\psi_{I}^{\rm br} \rangle$ of (\ref{new_a5}).
Examples of these states are shown in Table~\ref{tab3}, all located
around $\theta_1^{\rm lab} \approx 40^\circ$.

It appears that any combination of incoming polarizations can lead to
such entangled states, as exemplified at $E=135$~MeV in Table~\ref{tab3}.
Interestingly, all these states have practically vanishing
entanglement-spoiling components, whose magnitudes, in the worst cases,
reach only about $\approx 1 \%$ of the norm, as seen, for example, for
states $No~11$ and $12$ at $E=135$~MeV in Table~\ref{tab3}.

\begin{figure}
\includegraphics[scale=0.65]{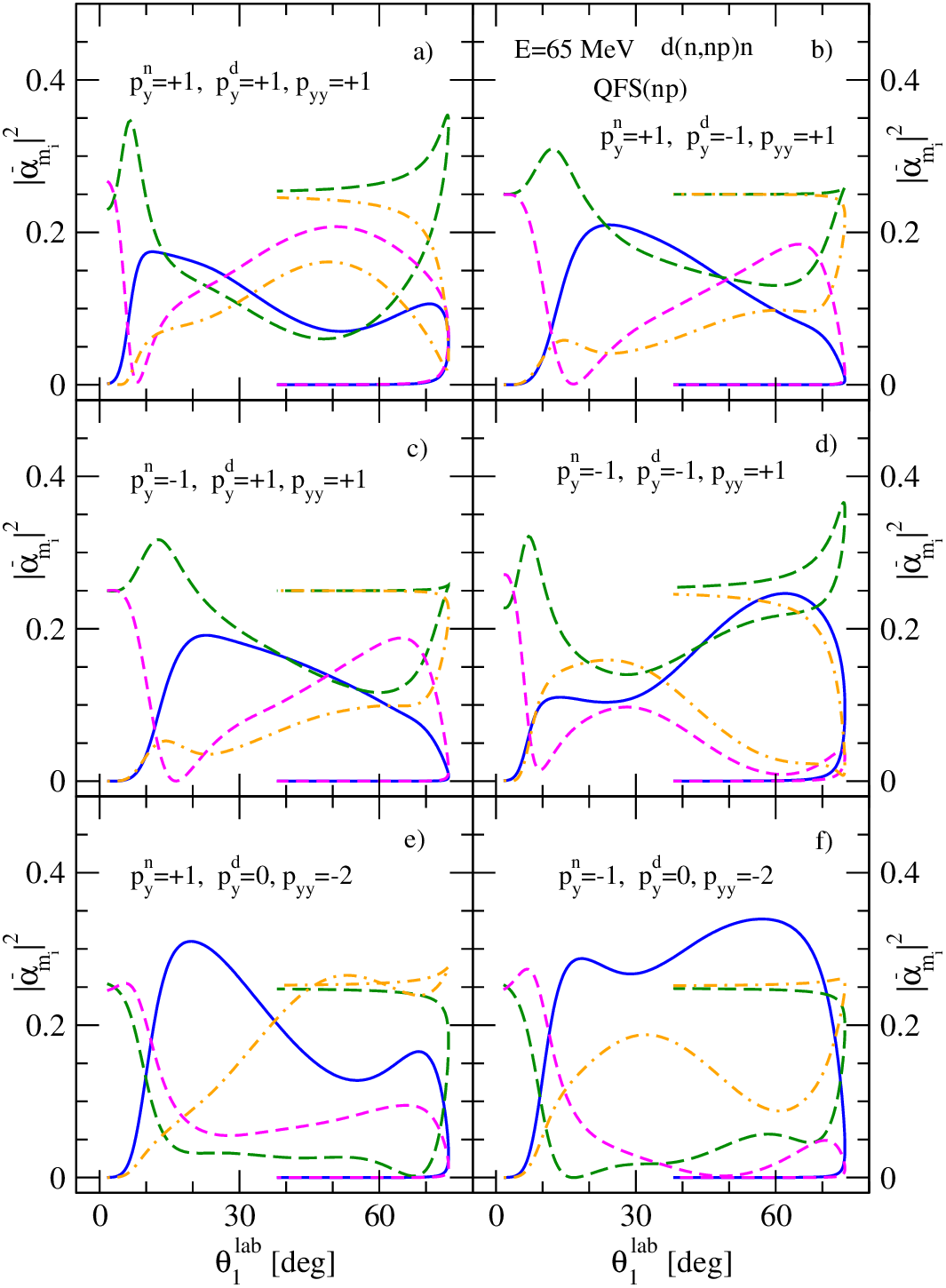}  
\caption{
  (color online)
 The same as in Fig.~\ref{fig2}, but for the QFS(np) case.
}
\label{fig3}
\end{figure}

The case of QFS($np$), exemplified at $E=65$~MeV in Fig.~\ref{fig3}, behaves
differently from QFS($nn$). A conspicuous lack of a clear approach to zero
for different subsets of lines corresponding to amplitudes with different
nucleon spin projections, together with their comparable magnitudes
over quite extended regions of $\theta_1^{\rm lab}$, makes it difficult
to locate entangled states with an admixture of the entanglement-spoiling
component below $\approx 10 \%$.

In addition, the large differences in magnitude of contributions shown
in Fig.~\ref{fig3}—either by the blue solid and green long-dashed lines
or by the magenta short-dashed and orange dash-dotted lines—are
disadvantageous for finding strongly entangled states.
This behavior of QFS($np$) with respect to QFS($nn$) can be traced
back to the different characteristics of free $np$ and $nn$ scattering,
where, in the former, a large contribution of the $s=1$ component is present.

Examples of entangled states found in this case are shown in Table~\ref{tab4}.

The most remarkable case is presented by the final-state-interaction 
geometry with two interacting neutrons, FSI($nn$), where, at each angle
$\theta_1^{\rm lab}$, the momenta of the two neutrons are equal:
$\vec{p}_{n_1} = \vec{p}_{n_2}$.
Here, all amplitudes with equal spin projections of these two neutrons
vanish, $\bar{\alpha}_{++m_3} = \bar{\alpha}_{--m_3} = 0$, while
the remaining amplitudes contribute equally to the norm of the final state:
$$|\bar{\alpha}_{+--}|^2=|\bar{\alpha}_{-++}|^2=|\bar{\alpha}_{+-+}|^2
=|\bar{\alpha}_{-+-}|^2= \frac {1} {4},$$
independently of $\theta_1^{\rm lab}$.

Consequently, for every combination of initial neutron and deuteron
polarizations, and at each angle $\theta_1^{\rm lab}$, one finds maximally
entangled Bell-like final spin states (\ref{new_a7}) of the $nn$ pair, without
any admixture of entanglement-spoiling components.
The exact vanishing of the $\bar{\alpha}_{++m_3}$ and $\bar{\alpha}_{--m_3}$
components results from the absolute dominance of the $^1S_0$ neutron–neutron
interaction in the FSI($nn$) geometry, which excludes the possibility of
equal spin projections for the two neutrons.

\begin{figure}
\includegraphics[scale=0.65]{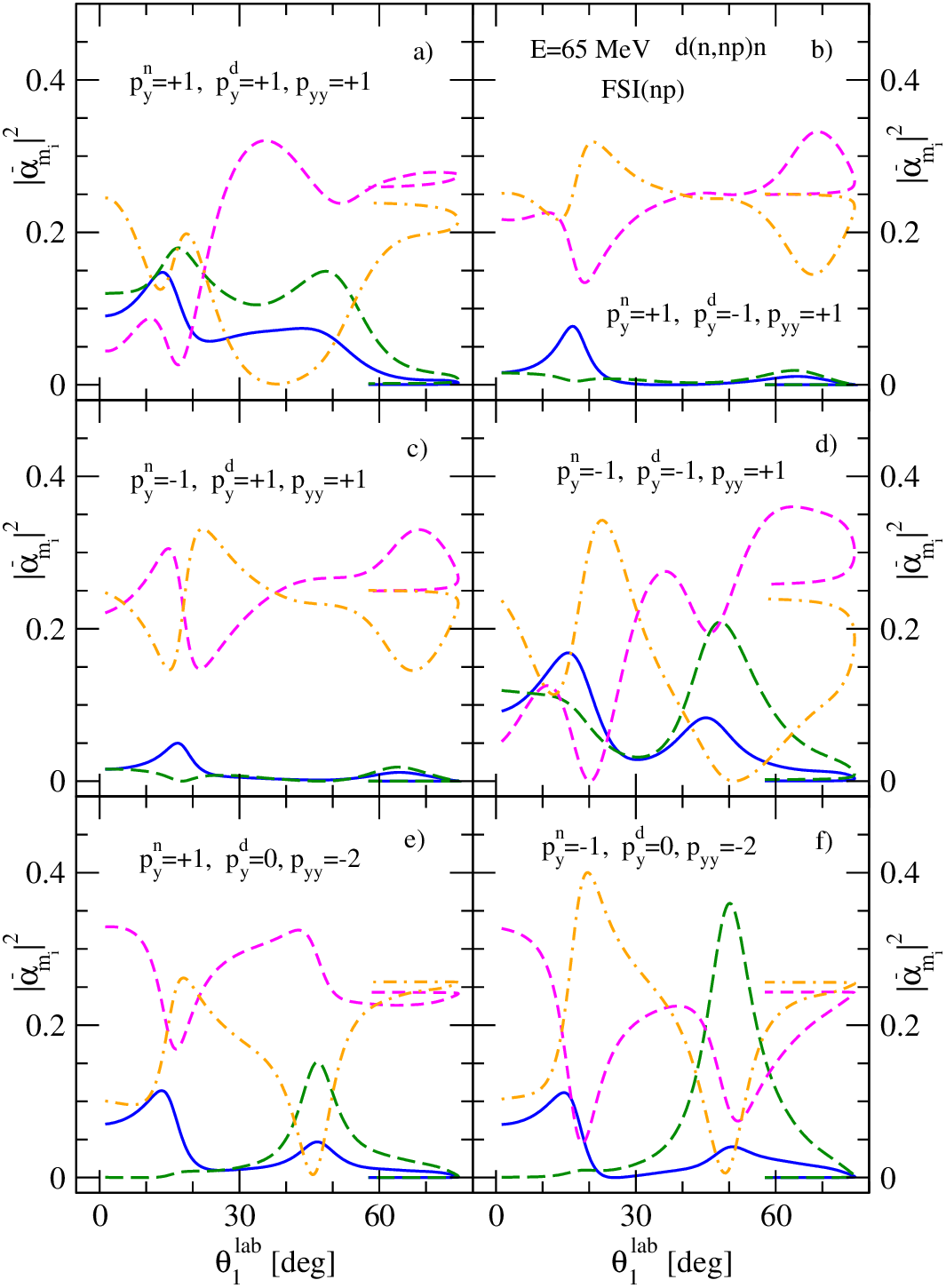}  
\caption{
  (color online)
  The same as in Fig.~\ref{fig2}, but for the FSI(np) case.
}
\label{fig4}
\end{figure}

Changing to the FSI($np$) geometry (exemplified at $E=65$~MeV in
Fig.~\ref{fig4}), where $\vec{p}_n = \vec{p}_p$ at each $\theta_1^{\rm lab}$,
one observes the behavior of different contributions to the norm that
is favorable for finding strongly entangled final $np$ spin states, as,
for example, in Figs.~\ref{fig4}b) and \ref{fig4}c).

Over quite a large region of the angle $\theta_1^{\rm lab}$, the contributions
shown by the blue solid lines
($|\bar{\alpha}_{---}|^2 = |\bar{\alpha}_{+++}|^2$) and the green long-dashed
lines ($|\bar{\alpha}_{--+}|^2 = |\bar{\alpha}_{++-}|^2$) vanish, while those
shown by the magenta short-dashed lines
($|\bar{\alpha}_{-+-}|^2 = |\bar{\alpha}_{+-+}|^2$) and orange dash-dotted
lines ($|\bar{\alpha}_{-++}|^2 = |\bar{\alpha}_{+--}|^2$) are of comparable
and large magnitudes.

In Table~\ref{tab5}, we present some of the entangled states found at the
four energies, located at forward angles. They are all Bell-like of
type (\ref{new_a5}), with only very small
contributions—below $\approx 1 \%$—from the entanglement-diminishing
component. These states are not restricted to forward angles only, but
can be found over wide regions of $\theta_1^{\rm lab}$.

The above results on the locations of entangled final states in the
deuteron breakup reaction were obtained by analyzing the magnitudes of
the different contributions to the norm.
When quantifying the degree of entanglement by considering the entanglement
power and concurrence, a natural question arises: would their application
yield similar results regarding the locations and the extent of the
$\theta_1^{\rm lab}$ ranges where final entangled states occur?

Before discussing the results, we note that, when defining the entanglement
power for the breakup according to Eq.~(\ref{a_eq1}), the reduced density
matrix $\rho_n$ must be obtained by tracing the final $8 \times 8$
breakup density matrix $\rho_{\rm out}$ over the subsystem of the two
remaining nucleons.

In the first step, we demonstrate the efficiency of applying the entanglement
power and concurrence to localize entangled states in kinematically complete
configurations. To this aim, we consider at $E=37.5$~MeV an exclusive breakup
$\vec d(\vec n,n_1 n_2)p$ and a complete configuration with
$\theta_1 = 45^\circ$, $\phi_1 = 0^\circ$,
$\theta_2 = 44.1^\circ$, $\phi_2 = 180^\circ$.
When viewed as a function of the arc length $S$, one finds that
at $S = 28$~MeV the exact condition for the QFS($nn$) geometry is
fulfilled, with the outgoing proton at rest, $\vec{p}_p = 0$,
and the first, $n_1$, and second, $n_2$, outgoing neutrons having momenta
directed along $\theta_1 = 45^\circ$, $\phi_1 = 0^\circ$ and
$\theta_2 = 44.1^\circ$, $\phi_2 = 180^\circ$, respectively.
This corresponds exactly to the QFS($nn$) geometry for which
the entangled states $No~3$ and $No~4$ of Table~\ref{tab3} were located.

\begin{figure}
\includegraphics[scale=0.63]{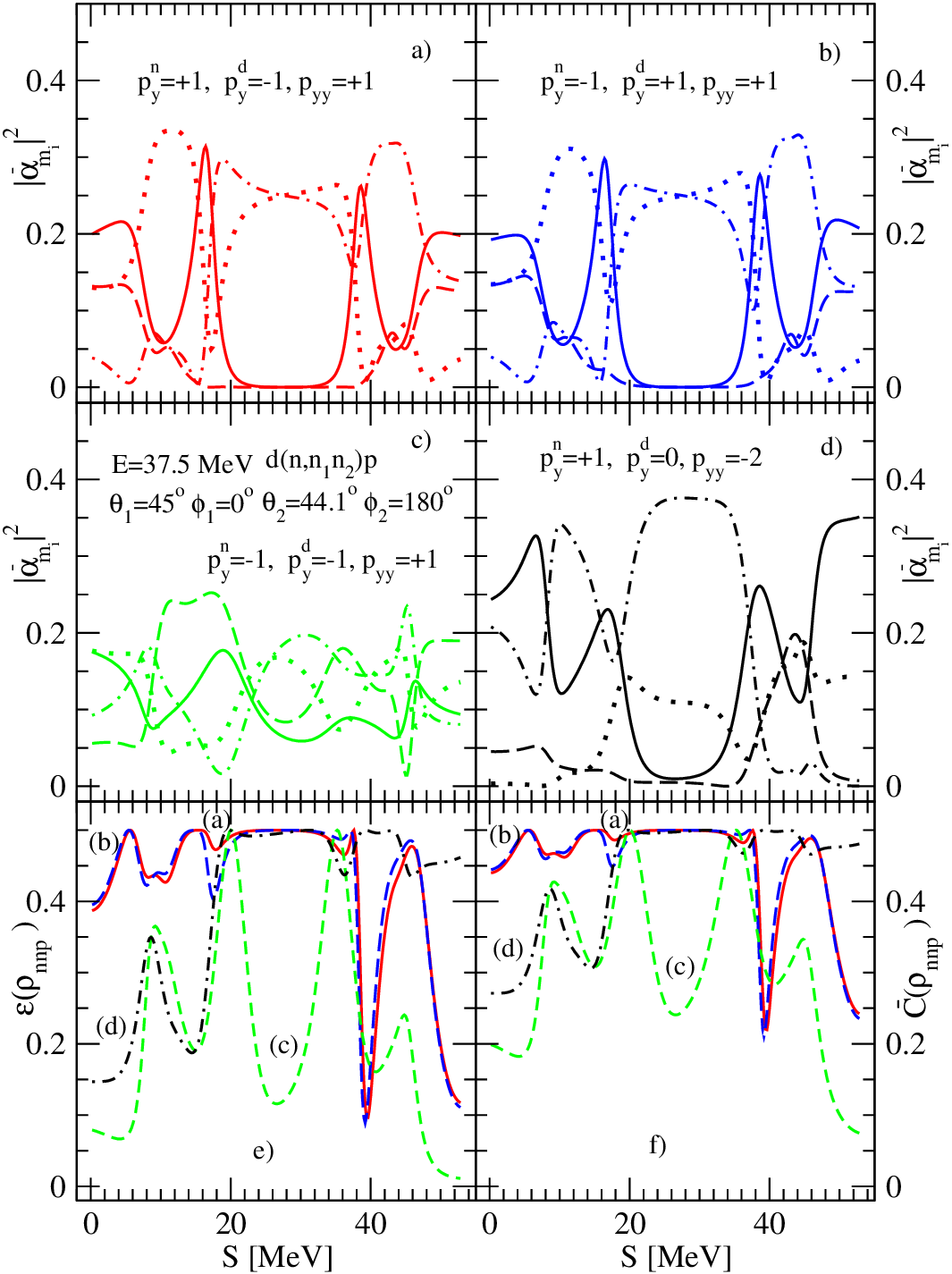}  
\caption{
  (color online)
  Distribution along the S-curve of contributions $|\bar{\alpha}_{m_1 m_2 m_3}|^2$
  to the norm for the final breakup states of Eq.~(\ref{new_a4}), formed
  in the QFS(nn) complete configuration of the $\vec d (\vec n, n_1 n_2)p$
  breakup reaction at $E = 37.5$~MeV and $\theta_1 = 45^\circ$,
  $\phi_1 = 0^\circ$, $\theta_2 = 44.1^\circ$, and $\phi_2 = 180^\circ$.
  The neutron and deuteron polarizations in pure cross-product initial states
  are shown in a)-d). 
  The solid, dashed, dotted, and dash-dotted lines in a)–d) show
  contributions of
$|\bar{\alpha}_{+++}|^2 = |\bar{\alpha}_{---}|^2$,
$|\bar{\alpha}_{++-}|^2 = |\bar{\alpha}_{--+}|^2$,
$|\bar{\alpha}_{+-+}|^2 = |\bar{\alpha}_{-+-}|^2$, and
$|\bar{\alpha}_{+--}|^2 = |\bar{\alpha}_{-++}|^2$, respectively.
  In e) and f), the distributions along the S-curve of the entanglement
  power $\epsilon(\rho_{nnp})$ and the concurrence ${\bar C}(\rho_{nnp})$
  for a)–d) are shown by the solid, short-dashed, long-dashed, and
  dash-dotted lines of the same color as in a), b), c), and d),
  respectively.
The predictions were obtained using the CD-Bonn potential.}
\label{fig5}
\end{figure}

In Figs.~\ref{fig5}a)–\ref{fig5}d), we show, for this particular complete
configuration, the contributions $|\bar{\alpha}_{m_1 m_2 m_3}|^2$ to the norm
for the breakup pure final spin states, formed from different combinations
of initial neutron and deuteron polarizations, as a function of the arc
length $S$. In Figs.~\ref{fig5}e) and \ref{fig5}f), the entanglement
power and concurrence corresponding to the final states shown
in panels (a)–(d) are displayed as functions of $S$. The color of each
curve in (e) and (f) matches that of the corresponding final state in (a)–(d).

The clear similarity between the behavior of the entanglement power and the
concurrence ensures that both measures lead to the same conclusions.
The maximal value of $0.5$ of the entanglement power at $S = 28$~MeV for
polarization combinations a), b), and d) supports the claim that entangled
states of the $nn$ pair exist at this location. The third state, for
polarization combination d), was not shown in Table~\ref{tab3} due to
larger contributions of entanglement-violating components, namely
$|\bar{\alpha}_{+++}|^2 = |\bar{\alpha}_{---}|^2 = 0.010$ and
$|\bar{\alpha}_{++-}|^2 = |\bar{\alpha}_{--+}|^2 = 0.005$.
It is interesting to note that the dominant components of this state
differ substantially and are equal to
$|\bar{\alpha}_{-+-}|^2 = |\bar{\alpha}_{+-+}|^2 = 0.109$ and
$|\bar{\alpha}_{-++}|^2 = |\bar{\alpha}_{+--}|^2 = 0.376$.

The flat maximum of the entanglement power (and also of the concurrence),
with a width comparable to that of the QFS peak itself, provides information
on how far within the QFS peak the entanglement is preserved. For
polarization combination (c) at this location, one finds a deep minimum
of the entanglement power. Indeed, the comparable magnitudes of the
different competing contributions in case (c) at $S = 28$~MeV exclude
the possibility of an entangled $nn$ state at this location.

It is noteworthy that, in case (c), the magnitudes of the contributions
indicate the presence of an entangled state at $S = 20$~MeV. This
interpretation is supported by a pronounced maximum of the entanglement
power at this location, with a value of approximately $0.5$
(see Fig.~\ref{fig5}e).

Establishing the efficiency of the entanglement power and concurrence in
locating entangled states along the S-curve, we apply them to study the
extent of the angular regions where QFS and FSI entangled states occur.

\begin{figure}
\includegraphics[scale=0.63]{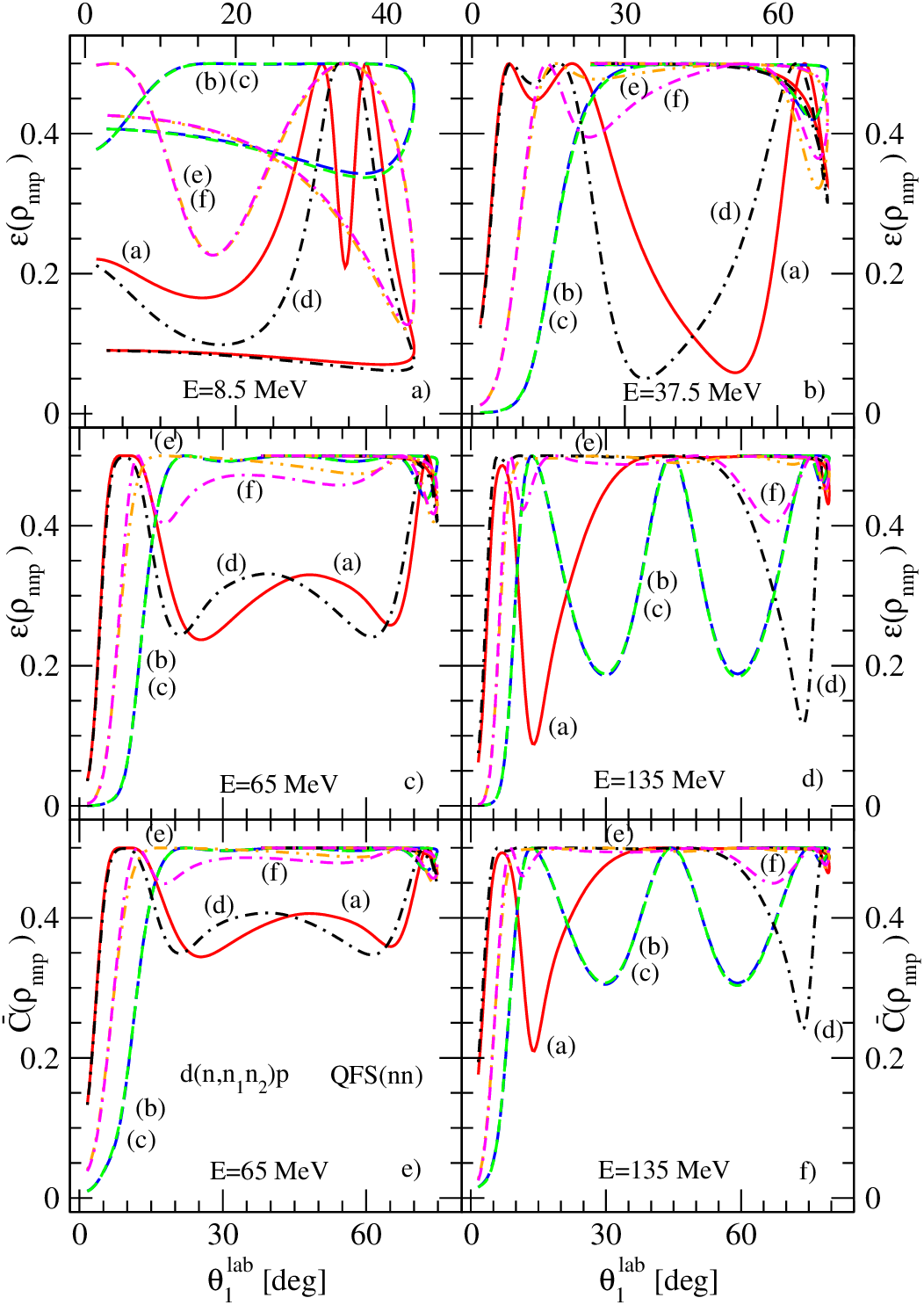}  
\caption{
  (color online)
  The entanglement power $\epsilon(\rho_{nnp})$ for the breakup pure final
  spin states of Eq.~(\ref{new_a4}) in the QFS(nn) exact geometry
  ($\vec p_p = 0$) at
  the laboratory angle of the first neutron $\theta_1^{lab}$.
  These states are formed in the $\vec d (\vec n, n_1 n_2)p$ breakup
  reaction at incoming neutron laboratory energies of $E = 8.5$~MeV a),
  $E = 37.5$~MeV b), $E = 65$~MeV c), and $E = 135$~MeV d), from
  pure cross-product initial states with the neutron and deuteron
  polarization combinations (a)–(f) given in Table~\ref{tab1}.
  The entanglement power $\epsilon(\rho_{nnp})$ is shown by different lines:
  (a) red solid, (b) blue long-dashed, (c) green short-dashed,
  (d) black dash-dotted, (e) orange dash-double-dotted, and
  (f) magenta double-dash-dotted.
  At $E = 65$~MeV e) and $E = 135$~MeV f), the concurrence
  ${\bar C}(\rho_{nnp})$ is also shown.
Predictions were obtained using the CD-Bonn potential. }
\label{fig6}
\end{figure}

In Figs.~\ref{fig6}a)–\ref{fig6}d), we present the entanglement power at
four energies and six polarization combinations of Table~\ref{tab1} for
the $\vec d(\vec n,n_1 n_2)p$ breakup in the QFS($nn$) geometry. At each
angle $\theta_1^{\rm lab}$, the QFS condition is fulfilled exactly.
In Figs.~\ref{fig6}e) and \ref{fig6}f), we show the corresponding
concurrence at $E=65$ and $135$~MeV, respectively. The contributions
$|\bar{\alpha}_{m_1 m_2 m_3}|^2$ to the norm at $E=65$~MeV are shown
in Fig.~\ref{fig2}.

We again find very similar behavior of the entanglement power and
concurrence, which ensures that the conclusions are independent of the
choice of measure. All states listed in Table~\ref{tab3} are rediscovered
using the entanglement power, and some of them belong to sets of
entangled states whose locations cover large regions of $\theta_1^{\rm lab}$
values. For example, states $No~3$ and $4$ at $E=37.5$~MeV and $No~5$
and $6$ at $E=65$~MeV are found within the range
$\theta_1^{\rm lab} \in (30^\circ, 55^\circ)$. Other states are well
localized, such as $No~1$ and $2$ at $E=8.5$~MeV, and $No~8$ and $9$
at $E=135$~MeV. In addition, many other entangled states, not included
in Table~\ref{tab3}, are found, many of which exhibit similarly
high entanglement quality as those shown.

\begin{figure}
\includegraphics[scale=0.65]{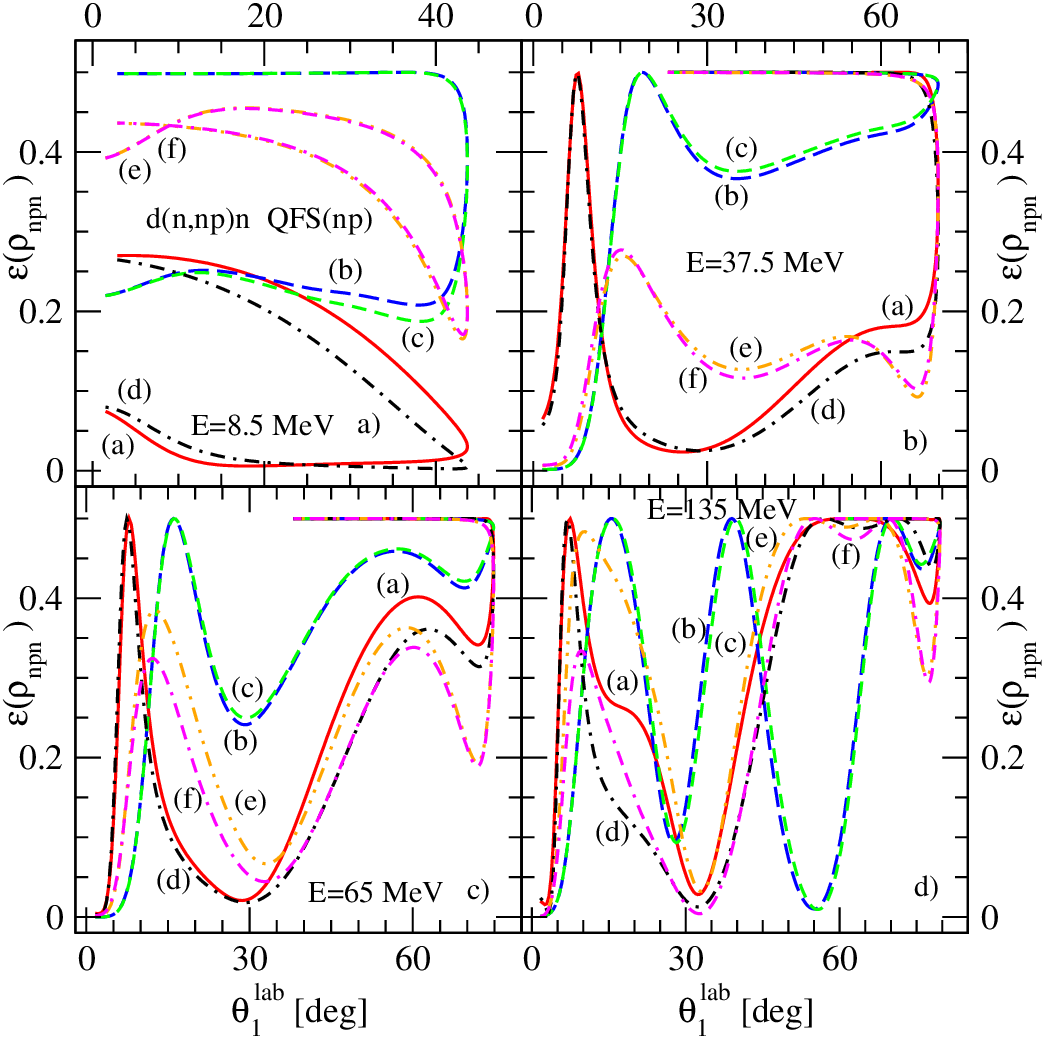}  
\caption{
  (color online)
  The same as in Fig.~\ref{fig6}, but for the breakup $\vec d (\vec n, np)n$
  pure final spin states of Eq.~(\ref{new_a4}) in the QFS(np)
  exact geometry ($\vec p_n = 0$).
}
\label{fig7}
\end{figure}

In the case of QFS($np$), we show in Figs.~\ref{fig7}a)–\ref{fig7}d) the
dependence of the entanglement power on $\theta_1^{\rm lab}$ at four energies.
At $E=8.5$~MeV (see Fig.~\ref{fig7}a)), $\epsilon (\rho_{npn})$ remains
below $0.5$ for all polarization combinations, supporting the absence
of entangled QFS $np$ pairs at this energy. Only for combinations (b)
and (c) does the entanglement power reach the value $0.5$ on the second
branch of solutions, which we excluded from the outset due to the very
small values of $\theta_2^{\rm lab}$ and $E_1^{\rm lab}$.

At energies $E=37.5$, $65$, and $135$~MeV, the entanglement power is again
small in most regions of $\theta_1^{\rm lab}$, with the exception of two
clear maxima reaching $0.5$ at forward angles observed at each energy.
These maxima correspond to the two pairs of polarization
combinations: ((a), (d)) at $\theta_1^{\rm lab} \approx 7.5^\circ$,
and ((b), (c)) at $\theta_1^{\rm lab} \approx 17.5^\circ$. For combination (d),
the entangled state at each energy has a contribution from
the entanglement-violating component of approximately $20 \%$ and was
therefore not included in Table~\ref{tab4}.

At $E=135$~MeV, six additional maxima are observed, all reaching $0.5$.
Two of them occur for the pair ((a), (d)) at
$\theta_1^{\rm lab} \approx 60^\circ$ and $70^\circ$, and two for
the pair ((b), (c)) at $\theta_1^{\rm lab} \approx 40^\circ$ and $70^\circ$.
At $\theta_1^{\rm lab} \approx 55^\circ$ and $70^\circ$, $\epsilon (\rho_{npn})$
approaches $0.5$ for the pair ((e), (f)). All of these maxima correspond
to entangled states, some with significant contributions
from entanglement-violating components.

We thus see that the entanglement power indeed reproduces all states
listed in Table~\ref{tab4}.

\begin{figure}
\includegraphics[scale=0.65]{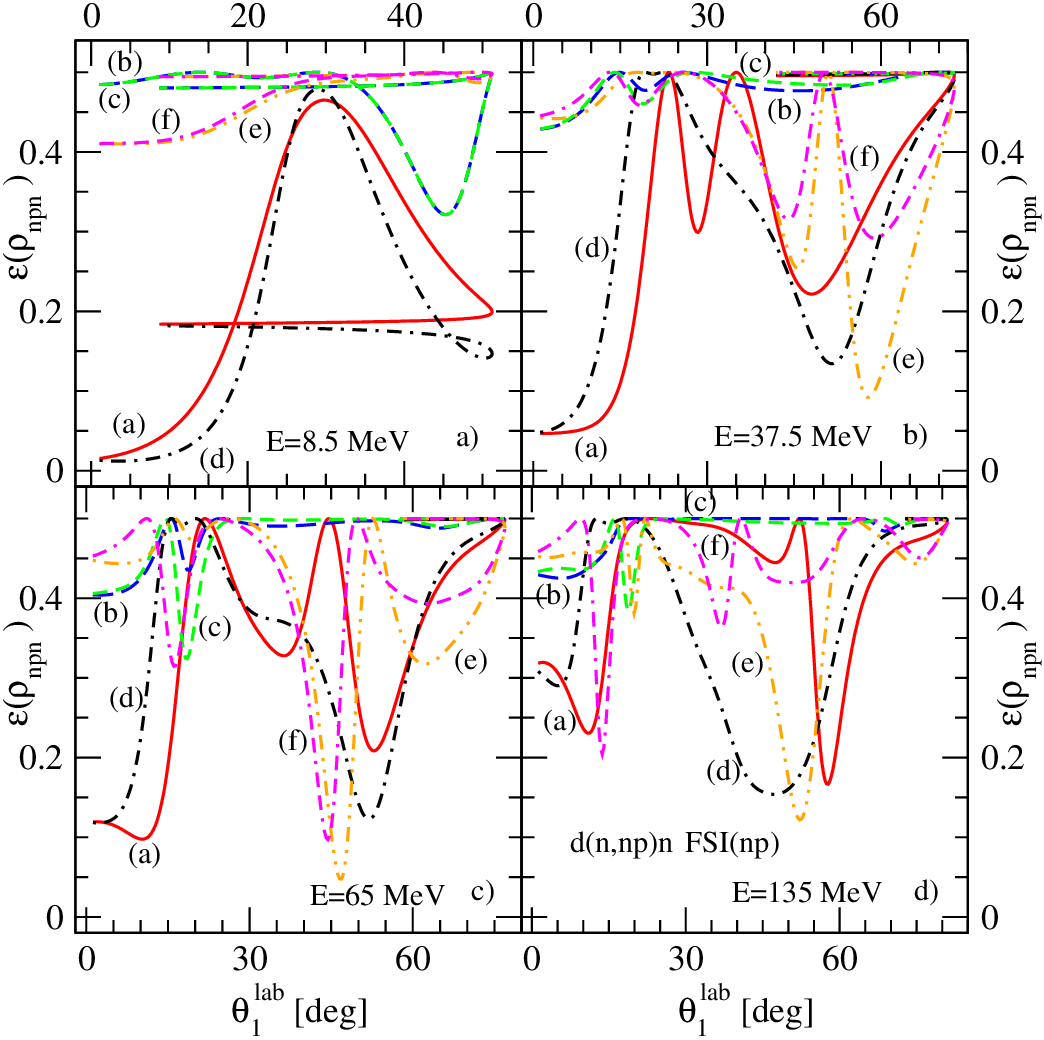}  
\caption{
  (color online)
  The same as in Fig.~\ref{fig6}, but for the breakup $\vec d (\vec n, np)n$
  pure final spin states of Eq.~(\ref{new_a4}) in the FSI(np) exact
  geometry ($\vec p_n = \vec p_p$).
}
\label{fig8}
\end{figure}

In Figs.~\ref{fig8}a)–\ref{fig8}d), we present the case of FSI($np$). The
entangled states listed in Table~\ref{tab5} are again rediscovered using
the entanglement power. For some combinations of incoming neutron and
deuteron polarizations, such as (a), (d), (e), and (f), the created
entangled states are localized within specific angular regions at each energy.
For combinations (b) and (c), the entangled states are formed over wide
angular regions, ranging, for example, at
$E=8.5$~MeV from $\theta_1^{\rm lab} \approx 10^\circ$ to $\approx 50^\circ$.

Finally, in the case of FSI($nn$) geometry, the corresponding figures
are very simple: the magnitudes of contributions to the norm from
different components at any energy are either $0$ or $\frac{1}{2}$, and
the entanglement power equals $\frac{1}{2}$ at all energies and angles
$\theta_1^{\rm lab}$ at which the FSI($n_1 n_2$) condition is fulfilled.
Consequently, all $nn$ pairs produced in the $\vec d(\vec n, n_1 n_2)p$
exclusive breakup in FSI($nn$) configurations are pure and maximally
entangled for all six incoming polarization combinations.

It is interesting to confirm the entanglement of all these states by a direct
calculation of the relevant spin observables, such as the outgoing
neutron polarization and the $N_1 N_2$ spin correlations.
A direct calculation gives:
\begin{eqnarray}
\langle \sigma_y^1 \otimes I^2  \otimes I^3 \rangle &\equiv& \langle \psi^{br} |
 \sigma_y^1 \otimes I^2  \otimes I^3 | \psi^{br} \rangle
   = \mp 4  Re [ \bar{\alpha}_{+++}^* \bar{\alpha}_{+--} -
   \bar{\alpha}_{++-}^* \bar{\alpha}_{+-+} ]  ~,
\label{eq_cor8}
\end{eqnarray}
and
\begin{eqnarray}
  \langle \sigma_y^1 \otimes \sigma_y^2  \otimes I^3
  \rangle &\equiv& \langle \psi^{br} |
 \sigma_y^1 \otimes \sigma_y^2  \otimes I^3 | \psi^{br} \rangle
   = \mp 4 Im [  \bar{\alpha}_{+++}  \bar{\alpha}_{++-}^* \mp 
   \bar{\alpha}_{+--} \bar{\alpha}_{+-+}^* ]  ~.
\label{eq_cor9}
\end{eqnarray}

For the final FSI($nn$) spin state, as expected, this gives a polarization
of the outgoing neutron $\langle \sigma_y^1 \rangle = 0$ and a spin
correlation $\langle \sigma_y^{n_1} \sigma_y^{n_2} \rangle = \mp 1$.

In Table~\ref{tab6}, we show the values of these observables for all states
listed in Tables~\ref{tab3}–\ref{tab5}. For states formed in QFS($nn$)
configurations, the numbers in Table~\ref{tab6} confirm nicely their
entanglement, with the exception of states $No~11$ and $12$. Most of
the states located in QFS($np$) geometries have either too large a
magnitude of the neutron polarization or too small a value of the spin
correlation. One can safely consider only state $No~7$ as a candidate
for a pure entangled Bell-like state in this geometry. For FSI($np$),
only state $No~1$ is clearly distinct from the others and should be
excluded as highly entangled.

\section{Summary and conclusions}
\label{sumary}

We investigated pure spin states of the outgoing neutron–deuteron pair
in elastic $nd$ scattering, as well as of the neutron–neutron and
neutron–proton pairs in the exclusive deuteron breakup reaction below
the pion production threshold, looking for signatures of their entanglement.
Such pure states can be created only from initial pure spin states of the
incoming neutron and deuteron, by polarizing them through maximal population
of one particular magnetic substate for each particle. In the case of axially
symmetric polarizations of the neutron and deuteron along the $y$-axis,
this allows for six combinations of polarized incoming pure states, each
of which was considered in this study.

By calculating at four energies the entanglement power and the concurrence
$\bar C$ for these final pure spin states, and by comparing the magnitudes
of contributions from different magnetic substates to their wave function,
we looked among them for strongly entangled states, with as small
an entanglement-spoiling component as possible.

We were not able to find such states in the case of elastic $nd$ scattering.
Admittedly, we found a few examples of entangled states at each of the
studied energies; however, these had rather large contributions from
entanglement-breaking admixtures. The purest entangled state we found,
at $E=135$~MeV, still contained approximately $10 \%$ of such admixture.

For the $nd$ exclusive breakup reaction, we looked for entangled pure
final spin states in two special complete geometries—QFS and FSI
configurations—which have the largest unpolarized cross sections.
We found in both very favorable conditions for the formation, from pure
incoming states of extremely polarized neutrons and deuterons, of strongly
entangled pure final spin states. Strong dominance of the singlet $^1S_0$
NN interaction in QFS($nn$) and FSI($np$) configurations allows the
formation of entangled final states with very small, practically vanishing,
admixtures of entanglement-spoiling components, over a wide range of
these geometries. The absolute dominance of the singlet $^1S_0$ NN
interaction in FSI($nn$) geometry causes the entanglement-spoiling
components to vanish entirely. Thus, any final state formed from extremely
polarized pure incoming neutron and deuteron states in this geometry
will be maximally entangled.

The question remains open whether presently used technology of polarized
ion sources allows the production of extremely polarized protons
and deuterons, with only one magnetic substate populated. If so, the
method of producing strongly entangled pairs of particles would be
available, opening interesting possibilities for investigations of
entangled states in the proton–deuteron breakup reaction.

It would also be interesting to explore, in a similar fashion, entanglement
in the final spin states of elastic scattering and breakup in four-nucleon
systems. Especially tempting would be elastic scattering and breakup in
the $\vec N + \vec {^3\rm He}$ and $\vec N + \vec {^3\rm H}$ systems,
which have spin structures identical to the nucleon–nucleon system.

\clearpage

\acknowledgements

This work was supported by the National Science Centre,
Poland under Grant
IMPRESS-U 2024/06/Y/ST2/00135.   
 It was also supported in part by the Excellence
Initiative – Research University Program at the Jagiellonian
University in Krak\'ow. 
The numerical calculations were partly performed on the supercomputers of
the JSC, J\"ulich, Germany.

\clearpage

%#------------------------------------------------------------------

\clearpage

%

%%%%%%%%%%%%%%%%%%%%%%%%%%%%%%%%%%

%\begin{sidewaystable}[htbp]
\begin{table}[ht!]
\centering
\caption{Values of the amplitudes, $\bar{\alpha}_{m_n m_d}$, and
  their contributions, $|\bar{\alpha}_{m_n m_d}|^2$, to the norm
  for the pure entangled spin states of Eq.~(\ref{new_a1}), at an
  incident neutron laboratory energy $E$ and center-of-mass
  angle $\Theta_{\mathrm{c.m.}}$.
  The polarization configuration of the initial state, as
  given in Table~\ref{tab1}, is indicated by the superscript
  on the energy value in the first column.
  The type of Bell state [as defined in Eqs.~(\ref{eq13_a0_a})
    and (\ref{eq13_a0_b})] is listed in the tenth column.
  Columns $A$, $B$, $C$, and $D$ give the average values of
  the correlations $\langle \sigma_y \otimes (P_y)^2 \rangle$,
  $\langle \sigma_y \otimes P_{yy} \rangle$, and the neutron
  and deuteron polarizations, $\langle \sigma_y \rangle$
  and $\langle P_y \rangle$, respectively. In columns 7, 8, and 9 we shortened
  notation for $\alpha$ leaving in index only $+$ or $-$ signs.
}
%\fontsize{9.0pt}{11pt}
\fontsize{8.0pt}{10pt}
\selectfont
\label{tab2}
\begin{tabular}{ |c|c|c|c|c|c|c|c|c|c|c|c|c|c|}
\hline
No & E$~^{x)}$  & $\Theta_{c.m.}$ & $\bar{\alpha}_{+\frac {1} {2} +1  }$
& $\bar{\alpha}_{+\frac {1} {2} 0} $
& $\bar{\alpha}_{+\frac {1} {2} -1} $
& $|\bar{\alpha}_{++}|^2$
& $|\bar{\alpha}_{+0}|^2$
& $|\bar{\alpha}_{+-}|^2$
&  Bell & $A$  & $B$ & $C$ & $D$ \\ 
& MeV & deg & $\bar{\alpha}_{-\frac {1} {2} -1}$
& $\bar{\alpha}_{-\frac {1} {2} 0  }$
& $\bar{\alpha}_{-\frac {1} {2} +1  }$  
& $|\bar{\alpha}_{--}|^2$  &
$|\bar{\alpha}_{-0}|^2$   &
$|\bar{\alpha}_{-+}|^2$ & type & &  & &\\
 \hline
1 & 8.5$~^{b)} $ &  $ 148 $ & $ -0.102 -i 0.579 $  & 
$ -0.227 +i 0.256 $ &  $ +0.139 -i 0.134 $  & 
$ 0.346 $  & $ 0.117 $   &$ 0.037 $  &  $| \psi_{I}^{nd} \rangle$
& $+0.490$ & $+1.510$ & $-0.019$ & $-0.460$ \\
 &   &   & $ -0.579 +i 0.102 $  & 
$ -0.256 -i 0.227 $ &   $ -0.134 -i 0.139 $  &
&    &   &  & & & & \\
\hline
2 & 8.5$~^{c)} $ &  $ 148 $ & $ +0.098 +i 0.613 $  & 
$ -0.144 +i 0.250 $ &  $ -0.145 +i 0.100 $  & 
$ 0.386 $  & $ 0.083 $   &$ 0.031 $  &  $| \psi_{I}^{nd} \rangle$
& $-0.488$ & $-1.512$ &  $+0.023$ & $+0.381$ \\
 &   &   & $ -0.613 +i 0.098 $  & 
$ +0.250 +i 0.144 $ &   $ -0.100 -i 0.145 $  &
&    &   &  & &  & & \\
\hline
3 & 37.5$~^{b)} $ &  $ 144 $ & $ +0.227 -i 0.612 $  & 
$ +0.230 +i 0.129 $ &  $ +0.051 -i 0.036 $  & 
$ 0.426 $  & $ 0.070 $   &$ 0.004 $  &  $| \psi_{I}^{nd} \rangle$
& $+0.503$ & $+1.497$ & $+0.005$ &  $+0.439$ \\
 &   &   & $ -0.612 -i 0.227 $  & 
$ -0.129 +i 0.230 $ &   $ -0.036 -i 0.051 $  &
&    &   &  & & & & \\
\hline
4 & 37.5$~^{c)} $ &  $ 147 $ & $ -0.224 +i 0.564 $  & 
$ +0.245 +i 0.232 $ &  $ -0.115 +i 0.1065 $  & 
$ 0.369 $  & $ 0.114 $   &$ 0.017 $  &  $| \psi_{I}^{nd} \rangle$
& $-0.489$  & $-1.511$ & $+0.022$ &  $-0.418$  \\
 &   &   & $ -0.564 -i 0.224 $  & 
$ +0.232 -i 0.245 $ &   $ -0.065 -i 0.115 $  &
&   &   &  & & & & \\
\hline
5 & 37.5$~^{f)} $ &  $ 148 $ & $ +0.296 -i 0.545 $  & 
$ -0.319 -i 0.069 $ &  $ +0.079 -i 0.053 $  & 
$ 0.384 $  & $ 0.106 $   &$ 0.009 $  &  $| \psi_{I}^{nd} \rangle$
& $+0.502$ & $+1.498$ & $+0.004$ & $-0.486$  \\
 &   &   & $ -0.548 -i 0.282 $  & 
$ +0.066 -i 0.328 $ &   $ -0.055 -i 0.073 $  &
&    &   &  & & &  & \\
\hline
6 & 65$~^{e)} $ &  $ 112 $ & $ +0.126 -i 0.216 $  & 
$ +0.025 -i 0.078 $ &  $ +0.579 +i 0.309 $  & 
$ 0.063 $  & $ 0.006 $   &$ 0.431 $  &  $| \psi_{II}^{nd} \rangle$
& $-0.494$ & $-1.506$ & $+0.012$ &  $+0.137$  \\
 &   &   & $ +0.216 +i 0.126 $  & 
$ -0.078 -i 0.025 $ &  $ -0.309 +i 0.579 $   & 
  &    &   &  & & & & \\
\hline
7 & 65$~^{f)} $ &  $ 146 $ & $ +0.499 -i 0.375 $  & 
$ -0.247 -i 0.009 $ &  $ +0.162 +i 0.150 $  & 
$ 0.390 $  & $ 0.061 $   &$ 0.049 $  &  $| \psi_{I}^{nd} \rangle$
& $+0.512$  & $+1.488$ & $+0.024$ &  $-0.376$  \\
 &   &   & $ -0.375 -i 0.499 $  & 
$ +0.009 -i 0.247 $ &   $ +0.150 -i 0.162 $  &
&    &   &  & & &  &  \\
\hline
8 & 135$~^{a)} $ & $ 88  $ & $ -0.063 +i 0.113 $  & $ +0.166 -i 0.006 $
& $ +0.542 +i 0.403 $
&  $ 0.017 $ & $ 0.027 $  &$ 0.456 $ &  $| \psi_{II}^{nd} \rangle$
& $+0.505$ &  $+1.495$ & $+0.010$ &  $+0.147$  \\
 &   &   & $ +0.113 +i 0.063 $  & 
$ +0.006 +i 0.166 $ &  $ +0.403 -i 0.542 $   & 
  &    &   &  & & & & \\
\hline
9 & 135$~^{e)} $ &  $83 $ & $ +0.093 +i 0.218 $  & 
$ -0.033 +i 0.311 $ &  $ +0.587 -i 0.034 $  & 
$ 0.056 $  & $ 0.098 $   & $ 0.346 $ &  $| \psi_{II}^{nd} \rangle$
& $-0.503$  & $-1.497$ & $-0.007$ & $-0.411$  \\
 &   &   & $ -0.218 +i 0.093 $  & 
$ +0.311 +i 0.033 $ &  $ +0.034 +i 0.587 $  & 
  &    &   & & &  & & \\
\hline
10 & 135$~^{e)} $ &  $109$ & $ +0.144 -i 0.215 $  & 
$ +0.192 +i 0.165 $ &  $ +0.577 +i 0.190 $  & 
$ 0.067 $  & $ 0.064 $   & $ 0.369 $ &  $| \psi_{II}^{nd} \rangle$
& $-0.500$  & $-1.500$ & $+0.001$ & $-0.384$ \\
 &   &   & $ -0.215 -i 0.144 $  & 
$ -0.165 +i 0.192 $ &  $ +0.190 -i 0.577 $  & 
  &    &   & & &  & & \\
\hline
11 & 135$~^{f)} $ &  $ 41 $ & $ -0.076 +i 0.124 $  & 
$ +0.298 -i 0.164 $ &  $ +0.231 +i 0.556 $  & 
$ 0.021 $  & $ 0.116 $   &$ 0.363 $  &  $| \psi_{II}^{nd} \rangle$
& $+0.513$  & $+1.487$ & $+0.026$ & $+0.507$  \\
 &   &   & $ +0.124 +i 0.076 $  & 
$ +0.164 +i 0.298 $ &  $ +0.556 -i 0.231 $   & 
  &    &   &  & & & & \\
\hline
12 & 135$~^{f)} $ &  $ 113 $ & $ +0.149 -i 0.170 $  & 
$ +0.210 +i 0.231 $ &  $ +0.556 +i 0.205 $  & 
$ 0.051 $  & $ 0.097 $   &$ 0.352 $  &  $| \psi_{II}^{nd} \rangle$
& $+0.501$ & $+1.499$ & $+0.003$ & $-0.444$   \\
 &   &   & $ -0.170 -i 0.149 $  & 
$ -0.231 +i 0.210 $ &   $ +0.205 -i 0.556 $  &
&    &   &  & & & & \\
\hline
\end{tabular}
\end{table}
%\end{sidewaystable}

%--------------------------------------------

%%%%%%%%%%%%%%%%%%%%%%%%%%%%%%%%%%

%\begin{sidewaystable}[htbp]
\begin{table}[ht!]
\centering
\caption{Values of the amplitudes $\bar{\alpha}_{m_1 m_2 m_3}$ and their
  contributions to the norm, $|\bar{\alpha}_{m_1 m_2 m_3}|^2$, for
  the deuteron breakup final pure spin states of Eq.~(\ref{new_a4})
  under the exact quasi-free scattering QFS(nn) condition.
  The incident neutron laboratory energy is $E$, and the laboratory
  production angle of the quasi-free scattered neutron–neutron pair
  is $\theta_{1}^{\mathrm{lab}}$. 
  These pure final states originate from pure initial states with
  the polarization combination $(p_y^n, p_y^d, p_{yy})$, indicated
  by the superscript on the energy value in the first column:
  a) $(+1,+1,+1)$, b) $(+1,-1,+1)$, c) $(-1,+1,+1)$,
  d) $(-1,-1,+1)$, e) $(+1,0,-2)$, and f) $(-1,0,-2)$.
}
\fontsize{9.0pt}{11pt}
\selectfont
\label{tab3}
\begin{tabular}{ |c|c|c|c|c|c|c|c|c|c|c|}
\hline
No & E$~^{a)}$  & $\theta_{1}^{\mathrm{lab}}$
& $\bar{\alpha}_{+++}$
& $\bar{\alpha}_{++-} $
& $\bar{\alpha}_{+-+} $
& $\bar{\alpha}_{+--} $
& $|\bar{\alpha}_{+++}|^2$
& $|\bar{\alpha}_{++-}|^2$
& $|\bar{\alpha}_{+-+}|^2$
& $|\bar{\alpha}_{+--}|^2$
  \\ 
& MeV & deg
& $\bar{\alpha}_{---}$
& $\bar{\alpha}_{--+}$
& $\bar{\alpha}_{-+-} $
& $\bar{\alpha}_{-++} $
& $|\bar{\alpha}_{---}|^2$  
& $|\bar{\alpha}_{--+}|^2$   
& $|\bar{\alpha}_{-+-}|^2$
& $|\bar{\alpha}_{-++}|^2$
  \\
\hline
1 & 8.5$~^{e)} $ & $ 35.0  $ & $ +0.008 +i 0.002 $  & $ -0.017 -i 0.006 $
& $ +0.408 +i 0.277 $ & $ -0.288 + i 0.416  $
&  $ 0.000 $ & $ 0.000 $  & $ 0.244 $ & 0.256      \\
&   &
& $ -0.002 +i 0.008 $  & $ -0.006 +i 0.017 $
&  $ +0.277 -i 0.408 $ & $ -0.416 - i 0.288 $
&    &   &  &   \\
\hline
2 & 8.5$~^{f)} $ & $ 35.0  $ & $ +0.015 -i 0.005 $  & $ +0.005 +i 0.000 $
& $ +0.418 +i 0.286 $ & $ +0.279 - i 0.406  $
&  $ 0.000 $ & $ 0.000 $  & $ 0.257 $ & 0.243      \\
&   &
& $ -0.005 -i 0.015 $  & $ -0.000 +i 0.005 $
&  $ -0.286 +i 0.418 $ & $ -0.406 - i 0.279 $
&    &   &  &   \\
\hline
3 & 37.5$~^{b)} $ & $ 45.0  $ & $ +0.006 -i 0.002 $  & $ -0.002 +i 0.006 $
& $ -0.327 -i 0.380 $ & $ -0.387 + i 0.315  $
&  $ 0.000 $ & $ 0.000 $  & $ 0.251 $ & 0.249      \\
&   &
& $ -0.002 -i 0.006 $  & $ -0.006 -i 0.002 $
&  $ +0.380 -i 0.327 $ & $ +0.315 + i 0.387 $
&    &   &  &   \\
\hline
4 & 37.5$~^{c)} $ & $ 45.0  $ & $ +0.000 +i 0.007 $  & $ +0.007 -i 0.014 $
& $ -0.320 -i 0.385 $ & $ +0.381 - i 0.322  $
&  $ 0.000 $ & $ 0.000 $  & $ 0.251 $ & 0.249     \\
&   &
& $ -0.007 +i 0.000 $  & $ -0.014 -i 0.007 $
&  $ -0.385 +i 0.320 $ & $ +0.322 + i 0.381 $
&    &   &  &   \\
\hline
5 & 65.0$~^{b)} $ & $ 43.9  $ & $ +0.002 +i 0.002 $  & $ -0.002 -i 0.004 $
& $ -0.365 -i 0.347 $ & $ -0.347 + i 0.355  $
&  $ 0.000 $ & $ 0.000 $  & $ 0.253 $ & 0.247    \\
&   &
& $ +0.002 -i 0.002 $  & $ -0.004 -i 0.002 $
&  $ +0.347 -i 0.365 $ & $ +0.355 + i 0.347 $
&    &   &  &   \\
\hline
6 & 65.0$~^{c)} $ & $ 43.9  $ & $ +0.002 +i 0.002 $  & $ +0.002 -i 0.004 $
& $ -0.355 -i 0.347 $ & $ +0.347 - i 0.365  $
&  $ 0.000 $ & $ 0.000 $  & $ 0.247 $ & 0.253    \\
&   &
& $ -0.002 +i 0.002 $  & $ -0.004 -i 0.002 $
&  $ -0.347 +i 0.355 $ & $ +0.365 + i 0.347 $
&    &   &  &   \\
\hline
7 & 135.0$~^{a)} $ & $ 48.0  $ & $ -0.000 -i 0.018 $  & $ -0.020 -i 0.007 $
& $ -0.069 +i 0.492 $ & $ -0.498 - i 0.068  $
&  $ 0.000 $ & $ 0.000 $  & $ 0.247 $ & 0.253    \\
&   &
& $ -0.018 +i 0.000 $  & $ +0.007 -i 0.020 $
&  $ -0.492 -i 0.069 $ & $ -0.068 + i 0.498 $
&    &   &  &   \\
\hline
8 & 135.0$~^{b)} $ & $ 44.6  $ & $ -0.006 -i 0.009 $  & $ +0.019 -i 0.017 $
& $ -0.418 -i 0.286 $ & $ -0.275 + i 0.409  $
&  $ 0.000 $ & $ 0.001 $  & $ 0.256 $ & 0.243     \\
&   &
& $ -0.009 +i 0.006 $  & $ +0.017 +i 0.019 $
&  $ +0.286 -i 0.418 $ & $ +0.409 + i 0.275 $
&    &   &  & \\
\hline
9 & 135.0$~^{c)} $ & $ 44.3  $ & $ -0.008 -i 0.007 $  & $ -0.020 +i 0.015 $
& $ -0.409 -i 0.275 $ & $ +0.286 - i 0.418  $
&  $ 0.000 $ & $ 0.001 $  & $ 0.243 $ & 0.256     \\
&   &
& $ +0.007 -i 0.008 $  & $ +0.015 +i 0.020 $
&  $ -0.275 +i 0.409 $ & $ +0.418 + i 0.286 $
&    &   &  &   \\
\hline
10 & 135.0$~^{d)} $ & $ 41.1  $ & $ +0.000 -i 0.017 $  & $ +0.021 +i 0.007 $
& $ +0.066 -i 0.498 $ & $ -0.492 - i 0.068  $
&  $ 0.000 $ & $ 0.000 $  & $ 0.254 $ & 0.246     \\
&   &
& $ +0.017 +i 0.000 $  & $ +0.007 -i 0.021 $
&  $ -0.498 -i 0.066 $ & $ +0.068 - i 0.492 $
&    &   &  &   \\
\hline
11 & 135.0$~^{e)} $ & $ 47.7  $ & $ +0.001 -i 0.002 $  & $ +0.051 +i 0.054 $
& $ +0.129 -i 0.396 $ & $ -0.567 + i 0.004  $
&  $ 0.000 $ & $ 0.005 $  & $ 0.174 $ & 0.321     \\
&   &
& $ +0.002 +i 0.001 $  & $ +0.054 -i 0.051 $
&  $ -0.396 -i 0.129 $ & $ -0.004 - i 0.567 $
&    &   &  &   \\
\hline
12 & 135.0$~^{f)} $ & $ 41.9  $ & $ -0.003 -i 0.004 $  & $ -0.046 -i 0.052 $
& $ +0.008 +i 0.567 $ & $ -0.398 - i 0.124  $
&  $ 0.000 $ & $ 0.005 $  & $ 0.322 $ & 0.173    \\
&   &
& $ -0.004 +i 0.003 $  & $ +0.052 -i 0.046 $
&  $ -0.567 +i 0.008 $ & $ -0.124 + i 0.398 $
&    &   &  &   \\
\hline
\end{tabular}
\end{table}
%\end{sidewaystable}

%--------------------------------------------

%--------------------------------------------

%%%%%%%%%%%%%%%%%%%%%%%%%%%%%%%%%%

%\begin{sidewaystable}[htbp]
\begin{table}[ht!]
\centering
\caption{Values of the amplitudes $\bar{\alpha}_{m_1 m_2 m_3}$ and their
  contributions to the norm, $|\bar{\alpha}_{m_1 m_2 m_3}|^2$, for the deuteron
  breakup final pure spin states of Eq.~(\ref{new_a4}) under the exact
  quasi-free scattering QFS(np) condition.
  The incident neutron laboratory energy is $E$, and the laboratory
  production angle of the quasi-free scattered neutron–proton pair
  is $\theta_{1}^{\mathrm{lab}}$.
  These pure final states originate from pure initial states with
  the polarization combination $(p_y^n, p_y^d, p_{yy})$, indicated
  by the superscript on the energy value in the first column:
  a) $(+1,+1,+1)$, b) $(+1,-1,+1)$, c) $(-1,+1,+1)$,
  d) $(-1,-1,+1)$, e) $(+1,0,-2)$, and f) $(-1,0,-2)$.
}
\fontsize{9.0pt}{11pt}
\selectfont
\label{tab4}
\begin{tabular}{ |c|c|c|c|c|c|c|c|c|c|c|}
\hline
No & E$~^{a)}$  & $\theta_{1}^{\mathrm{lab}}$
& $\bar{\alpha}_{+++}$
& $\bar{\alpha}_{++-} $
& $\bar{\alpha}_{+-+} $
& $\bar{\alpha}_{+--} $
& $|\bar{\alpha}_{+++}|^2$
& $|\bar{\alpha}_{++-}|^2$
& $|{\bar\alpha}_{+-+}|^2$
& $|\bar{\alpha}_{+--}|^2$
  \\ 
& MeV & deg
& ${\bar\alpha}_{---}$
& $\bar{\alpha}_{--+}$
& ${\bar\alpha}_{-+-} $
& $\bar{\alpha}_{-++} $
& $|\bar{\alpha}_{---}|^2$  
& $|\bar{\alpha}_{--+}|^2$   
& $|\bar{\alpha}_{-+-}|^2$
& $|\bar{\alpha}_{-++}|^2$
  \\
\hline
1 & 37.5$~^{a)} $ & $ 8.6  $ & $ -0.380 +i 0.008 $  & $ -0.017 -i 0.525 $
& $ +0.170 -i 0.042 $ & $ +0.179 + i 0.132  $
&  $ 0.144 $ & $ 0.276 $  & $ 0.030 $ & 0.050      \\
&   &
& $ +0.008 +i 0.380 $  & $ +0.525 -i 0.017 $
&  $ +0.042 +i 0.170 $ & $ +0.132 - i 0.179 $
&    &   &  &   \\
\hline
2 & 37.5$~^{b)} $ & $ 17.5  $ & $ -0.406 -i 0.012 $  & $ -0.286 +i 0.437 $
& $ +0.217 +i 0.025 $ & $ +0.030 + i 0.120  $
&  $ 0.165 $ & $ 0.272 $  & $ 0.048 $ & 0.015      \\
&   &
& $ -0.012 +i 0.406 $  & $ -0.437 -i 0.286 $
&  $ -0.025 +i 0.217 $ & $ +0.120 - i 0.030 $
&    &   &  &   \\
\hline
3 & 37.5$~^{c)} $ & $ 17.5  $ & $ +0.400 -i 0.009 $  & $ -0.309 +i 0.436 $
& $ +0.199 +i 0.030 $ & $ -0.039 - i 0.107  $
&  $ 0.160 $ & $ 0.286 $  & $ 0.041 $ & 0.013      \\
&   &
& $ +0.009 +i 0.400 $  & $ +0.436 +i 0.309 $
&  $ +0.030 -i 0.199 $ & $ +0.107 - i 0.039 $
&    &   &  &   \\
\hline
4 & 65.0$~^{a)} $ & $ 8.3  $ &
$ -0.333 +i 0.210 $  & $ -0.126 -i 0.538 $
& $ -0.057 +i 0.012 $ & $ +0.190 - i 0.006  $
&  $ 0.155 $ & $ 0.305 $  & $ 0.004 $ & 0.036     \\
&   &
& $ +0.210 +i 0.333 $  & $ +0.538 -i 0.126 $
&  $ -0.012 -i 0.057 $ & $ -0.006 - i 0.190 $
&    &   &  &   \\
\hline
5 & 65.0$~^{b)} $ & $ 16.4  $ & $ -0.395 +i 0.110 $  & $ -0.063 +i 0.520 $
& $ +0.030 -i 0.014 $ & $ +0.105 + i 0.212  $
&  $ 0.168 $ & $ 0.275 $  & $ 0.001 $ & 0.056    \\
&   &
& $ +0.110 +i 0.395 $  & $ -0.520 -i 0.063 $
&  $ +0.014 +i 0.030 $ & $ +0.212 - i 0.105 $
&    &   &  &   \\
\hline
6 & 65.0$~^{c)} $ & $ 16.4  $ & $ +0.377 -i 0.131 $  & $ -0.091 +i 0.532 $
& $ +0.016 +i 0.001 $ & $ -0.106 - i 0.195  $
&  $ 0.159 $ & $ 0.291 $  & $ 0.000 $ & 0.050    \\
&   &
& $ +0.131 +i 0.377 $  & $ +0.532 +i 0.091 $
& $ +0.001 -i 0.016 $ & $ +0.195 - i 0.106 $
&    &   &  &   \\
\hline
7 & 135.0$~^{b)} $ & $ 40.8  $ & $ -0.052 -i 0.480 $  & $ -0.461 +i 0.068 $
& $ +0.055 +i 0.131 $ & $ +0.161 - i 0.056  $
&  $ 0.233 $ & $ 0.218 $  & $ 0.020 $ & 0.029    \\
&   &
& $ -0.480 +i 0.052 $  & $ -0.068 -i 0.461 $
&  $ -0.131 +i 0.055 $ & $ -0.056 - i 0.161 $
&    &   &  &   \\
\hline
8 & 135.0$~^{b)} $ & $ 67.7  $ & $ -0.068 -i 0.057 $  & $ -0.088 +i 0.188 $
& $ +0.496 -i 0.010 $ & $ +0.034 - i 0.449  $
&  $ 0.008 $ & $ 0.043 $  & $ 0.246 $ & 0.203     \\
&   &
& $ -0.057 +i 0.068 $  & $ -0.188 -i 0.088 $
&  $ +0.010 +i 0.496 $ & $ -0.068 - i 0.057 $
&    &   &  & \\
\hline
9 & 135.0$~^{c)} $ & $ 40.8  $ & $ +0.041 +i 0.479 $  & $ -0.456 +i 0.062 $
& $ +0.059 +i 0.136 $ & $ -0.164 + i 0.093  $
&  $ 0.231 $ & $ 0.212 $  & $ 0.022 $ & 0.035     \\
&   &
& $ -0.479 +i 0.041 $  & $ +0.062 +i 0.456 $
&  $ +0.136 -i 0.059 $ & $ -0.093 - i 0.164 $
&    &   &  &   \\
\hline
10 & 135.0$~^{c)} $ & $ 67.2  $ & $ +0.102 +i 0.085 $  & $ -0.080 +i 0.155 $
& $ +0.506 -i 0.024 $ & $ -0.058 + i 0.438  $
&  $ 0.018 $ & $ 0.030 $  & $ 0.257 $ & 0.195     \\
&   &
& $ -0.085 +i 0.102 $  & $ +0.155 +i 0.080 $
&  $ -0.024 -i 0.506 $ & $ -0.438 - i 0.058 $
&    &   &  &   \\
\hline
\end{tabular}
\end{table}
%\end{sidewaystable}

%--------------------------------------------

%\begin{sidewaystable}[htbp]
\begin{table}[ht!]
\centering
\caption{Values of the amplitudes $\bar{\alpha}_{m_1 m_2 m_3}$ and their
  contributions to the norm, $|\bar{\alpha}_{m_1 m_2 m_3}|^2$, for the deuteron
  breakup final pure spin states of Eq.~(\ref{new_a4}) under the exact
  final-state interaction FSI(np) condition.
  The incident neutron laboratory energy is $E$, and the laboratory
  production angle of the final-state interacting neutron–proton pair
  is $\theta_{1}^{\mathrm{lab}}$.
  These pure final states originate from pure initial states with
  the polarization combination $(p_y^n, p_y^d, p_{yy})$, indicated
  by the superscript on the energy value in the first column:
  a) $(+1,+1,+1)$, b) $(+1,-1,+1)$, c) $(-1,+1,+1)$,
  d) $(-1,-1,+1)$, e) $(+1,0,-2)$, and f) $(-1,0,-2)$.
}
\fontsize{9.0pt}{11pt}
\selectfont
\label{tab5}
\begin{tabular}{ |c|c|c|c|c|c|c|c|c|c|c|}
\hline
No & E$~^{a)}$  & $\theta_{1}^{\mathrm{lab}}$
& ${\alpha}_{+++}$
& ${\alpha}_{++-} $
& ${\alpha}_{+-+} $
& ${\alpha}_{+--} $
& $|{\alpha}_{+++}|^2$
& $|{\alpha}_{++-}|^2$
& $|{\alpha}_{+-+}|^2$
& $|{\alpha}_{+--}|^2$
  \\ 
& MeV & deg
& ${\alpha}_{---}$
& ${\alpha}_{--+}$
& ${\alpha}_{-+-} $
& ${\alpha}_{-++} $
& $|{\alpha}_{---}|^2$  
& $|{\alpha}_{--+}|^2$   
& $|{\alpha}_{-+-}|^2$
& $|{\alpha}_{-++}|^2$
  \\
\hline
1 & 8.5$~^{a)} $ & $ 29.6  $ & $ -0.090 -i 0.014 $  & $ +0.036 +i 0.542 $
& $ -0.253 +i 0.081 $ & $ +0.614 + i 0.199  $
&  $ 0.008 $ & $ 0.004 $  & $ 0.071 $ & 0.417      \\
&   &
& $ -0.014 +i 0.090 $  & $ -0.054 +i 0.036 $
&  $ -0.081 -i 0.253 $ & $ +0.199 - i 0.614 $
&    &   &  &   \\
\hline
2 & 8.5$~^{b)} $ & $ 18.0  $ & $ -0.007 -i 0.084 $  & $ +0.033 -i 0.090 $
& $ +0.118 -i 0.416 $ & $ -0.509 - i 0.193  $
&  $ 0.007 $ & $ 0.009 $  & $ 0.187 $ & 0.297      \\
&   &
& $ -0.084 +i 0.007 $  & $ +0.090 +i 0.033 $
&  $ +0.416 +i 0.118 $ & $ -0.193 + i 0.509 $
&    &   &  &   \\
\hline
3 & 8.5$~^{c)} $ & $ 18.0  $ & $ +0.000 +i 0.079 $  & $ +0.030 -i 0.084 $
& $ +0.121 -i 0.405 $ & $ +0.513 + i 0.210  $
&  $ 0.006 $ & $ 0.008 $  & $ 0.179 $ & 0.307      \\
&   &
& $ -0.079 +i 0.000 $  & $ -0.084 -i 0.030 $
&  $ -0.405 -i 0.121 $ & $ -0.210 + i 0.513 $
&    &   &  &   \\
\hline
4 & 37.5$~^{b)} $ & $ 39.0  $ & $ -0.032 +i 0.029 $  & $ +0.057 -i 0.071 $
& $ -0.466 -i 0.114 $ & $ -0.113 + i 0.497  $
&  $ 0.002 $ & $ 0.008 $  & $ 0.230 $ & 0.260     \\
&   &
& $ +0.029 +i 0.032 $  & $ +0.071 +i 0.057 $
&  $ +0.114 -i 0.466 $ & $ +0.497 + i 0.113 $
&    &   &  &   \\
\hline
5 & 37.5$~^{c)} $ & $ 39.0  $ & $ -0.005 -i 0.057 $  & $ +0.052 -i 0.018 $
& $ -0.493 -i 0.053 $ & $ +0.164 - i 0.470  $
&  $ 0.003 $ & $ 0.003 $  & $ 0.246 $ & 0.248     \\
&   &
& $ +0.057 -i 0.005 $  & $ -0.018 -i 0.052 $
&  $ -0.053 +i 0.493 $ & $ +0.470 + i 0.164 $
&    &   &  &   \\
\hline
6 & 37.5$~^{e)} $ & $ 25.1  $ & $ -0.045 -i 0.023 $  & $ +0.010 -i 0.010 $
& $ +0.379 +i 0.322 $ & $ -0.403 + i 0.297  $
&  $ 0.003 $ & $ 0.000 $  & $ 0.247 $ & 0.250     \\
&   &
& $ +0.023 -i 0.045 $  & $ -0.010 -i 0.010 $
&  $ +0.322 -i 0.379 $ & $ -0.297 - i 0.403 $
&    &   &  &   \\
\hline
7 & 65.0$~^{b)} $ & $ 37.0  $ & $ -0.002 +i 0.002 $  & $ +0.044 -i 0.042 $
& $ -0.469 -i 0.148 $ & $ -0.048 + i 0.502  $
&  $ 0.000 $ & $ 0.003 $  & $ 0.242 $ & 0.255    \\
&   &
& $ +0.002 +i 0.002 $  & $ +0.042 +i 0.044 $
&  $ +0.148 -i 0.469 $ & $ +0.502 + i 0.048 $
&    &   &  &   \\
\hline
8 & 65.0$~^{c)} $ & $ 47.1  $ & $ -0.003 -i 0.040 $  & $ +0.007 -i 0.003 $
& $ -0.506 +i 0.098 $ & $ -0.022 - i 0.482  $
&  $ 0.001 $ & $ 0.000 $  & $ 0.266 $ & 0.233    \\
&   &
& $ +0.040 -i 0.003 $  & $ -0.003 -i 0.007 $
&  $ +0.098 +i 0.506 $ & $ +0.482 - i 0.022 $
&    &   &  &   \\
\hline
9 & 65.0$~^{e)} $ & $ 26.0  $ & $ -0.076 -i 0.061 $  & $ +0.078 -i 0.060 $
& $ +0.455 +i 0.265 $ & $ -0.303 + i 0.334  $
&  $ 0.009 $ & $ 0.010 $  & $ 0.278 $ & 0.203    \\
&   &
& $ +0.061 -i 0.076 $  & $ -0.060 -i 0.078 $
&  $ +0.265 -i 0.455 $ & $ -0.334 - i 0.303 $
&    &   &  &   \\
\hline
10 & 65.0$~^{f)} $ & $ 24.4  $ & $ -0.001 -i 0.020 $  & $ -0.081 +i 0.058 $
& $ +0.304 +i 0.249 $ & $ +0.468 - i 0.340  $
&  $ 0.000 $ & $ 0.010 $  & $ 0.155 $ & 0.335    \\
&   &
& $ -0.020 +i 0.001 $  & $ -0.058 -i 0.081 $
&  $ -0.249 +i 0.304 $ & $ -0.340 - i 0.468 $
&    &   &  &   \\
\hline
11 & 135.0$~^{b)} $ & $ 59.0  $ & $ +0.003 -i 0.029 $  & $ -0.031 -i 0.002 $
& $ -0.577 +i 0.146 $ & $ +0.133 + i 0.355  $
&  $ 0.001 $ & $ 0.001 $  & $ 0.354 $ & 0.144    \\
&   &
& $ -0.029 -i 0.003 $  & $ +0.002 -i 0.031 $
&  $ -0.146 -i 0.577 $ & $ +0.355 - i 0.133 $
&    &   &  &   \\
\hline
12 & 135.0$~^{c)} $ & $ 59.0  $ & $ -0.030 +i 0.068 $  & $ +0.050 +i 0.008 $
& $ -0.558 +i 0.142 $ & $ -0.126 - i 0.380  $
&  $ 0.006 $ & $ 0.002 $  & $ 0.332 $ & 0.160     \\
&   &
& $ -0.068 -i 0.030 $  & $ +0.008 -i 0.050 $
&  $ +0.142 +i 0.558 $ & $ +0.380 - i 0.126 $
&    &   &  & \\
\hline
13 & 135.0$~^{d)} $ & $ 36.4  $ & $ -0.000 -i 0.001 $  & $ +0.083 +i 0.016 $
& $ -0.472 +i 0.109 $ & $ +0.069 - i 0.503  $
&  $ 0.000 $ & $ 0.007 $  & $ 0.235 $ & 0.258     \\
&   &
& $ +0.001 -i 0.000 $  & $ +0.016 -i 0.083 $
&  $ +0.109 +i 0.472 $ & $ +0.503 + i 0.069 $
&    &   &  &   \\
\hline
14 & 135.0$~^{e)} $ & $ 66.0  $ & $ +0.002 +i 0.006 $  & $ +0.069 -i 0.051 $
& $ -0.387 -i 0.287 $ & $ +0.231 - i 0.455  $
&  $ 0.000 $ & $ 0.007 $  & $ 0.233 $ & 0.260     \\
&   &
& $ -0.006 +i 0.002 $  & $ -0.051 -i 0.069 $
&  $ -0.287 +i 0.387 $ & $ +0.455 + i 0.231 $
&    &   &  &   \\
\hline
15 & 135.0$~^{f)} $ & $ 78.2  $ & $ -0.073 +i 0.049 $  & $ +0.070 +i 0.065 $
& $ -0.475 -i 0.079 $ & $ -0.087 + i 0.494  $
&  $ 0.008 $ & $ 0.009 $  & $ 0.232 $ & 0.251     \\
&   &
& $ +0.049 +i 0.073 $  & $ -0.065 +i 0.070 $
&  $ +0.079 -i 0.475 $ & $ +0.494 + i 0.087 $
&    &   &  &   \\
\hline
\end{tabular}
\end{table}
%\end{sidewaystable}

%

%%%%%%%%%%%%%%%%%%%%%%%%%%%%%%%%%%

%\begin{sidewaystable}[htbp]
\begin{table}[ht!]
\centering
\caption{Values of the nucleon polarization, $\langle \sigma_y^{N_1} \rangle$,
  and the nucleon–nucleon spin correlation,
  $\langle \sigma_y^{N_1} \otimes \sigma_y^{N_2} \rangle$, for all
  entangled states listed in Tables~\ref{tab3}–\ref{tab5}, formed under
  the quasi-free scattering QFS($N_1N_2$) and final-state
  interaction FSI($np$) kinematically complete geometries of
  the exclusive breakup reaction $\vec{d}(\vec{n}, N_1 N_2) N_3$.
}
\fontsize{9.0pt}{11pt}
\selectfont
\label{tab6}
\begin{tabular}{ |c|c|c|c|c|c|c|}
\hline
No & QFS(nn)  & QFS(nn) & QFS(np) & QFS(np) & FSI(np) & FSI(np) \\
  & $\langle \sigma_y^{n} \rangle$ & $\langle \sigma_y^n \otimes \sigma_y^n
\rangle$  &
   $\langle \sigma_y^{n} \rangle$ & $\langle \sigma_y^n \otimes \sigma_y^p
   \rangle$  &
   $\langle \sigma_y^{n} \rangle$ & $\langle \sigma_y^n \otimes \sigma_y^p
   \rangle$  \\
 \hline
1 & $+0.014$ &  $ +0.999 $ & $+0.345$  & 
$+0.917$ &  $+0.213$  &  $-0.418$   \\
\hline
2 & $-0.002$ &  $-0.999$ & $-0.149$  & 
$-0.622$ &  $+0.087$  &  $-0.925$   \\
\hline
3 & $+0.005$ &  $-0.999$ & $+0.136$  & 
$-0.768$ &  $-0.083$  &  $+0.942$   \\
\hline
4 & $-0.020$ &  $+0.999$ & $+0.262$  & 
$+0.815$ &  $-0.145$  &  $-0.976$    \\
\hline
5 & $-0.000$ &  $-1.000$ & $+0.036$  & 
$-0.763$ &  $+0.202$  &  $+0.950$   \\
\hline
6 & $-0.002$ &  $+1.000$ & $-0.053$  & 
$-0.767$ &  $+0.043$  &  $+0.965$   \\
\hline
7 & $-0.014$ &  $+0.997$ & $-0.141$  & 
$-0.997$ &  $-0.062$  &  $-0.970$   \\
\hline
8 & $-0.003$ &  $-0.997$ & $-0.276$  & 
$-0.961$ &  $+0.093$  & $+0.983$   \\
\hline
9 & $-0.014$ &  $+0.999$ & $+0.224$  & 
$-0.773$ &  $-0.067$  &   $+0.892$   \\
\hline
10 & $+0.012$ &  $-1.000$ & $+0.302$  & 
$+0.790$ &  $-0.066$  &  $-0.888$   \\
\hline
11 & $+0.057$ & $-0.896$ &  &  & $+0.111$ & $-0.901$  \\
\hline
12 & $-0.125$ &  $+0.898$ &   &  & $+0.017$   & $+0.935$   \\
\hline
13 &  &   &  &  & $+0.152$  & $+0.920$  \\
\hline
14 &  &   &   &   & $+0.038$  & $+0.972$  \\
\hline
15 &  &  &   &   & $-0.276$   & $-0.998$   \\
\hline
\end{tabular}
\end{table}
%\end{sidewaystable}

%--------------------------------------------

%%%%%%%%%%%%%%%%%%%%%%%%%%%%%%%%%%

%--------------------------------------------

\clearpage

\end{document}